# A closer physico-chemical look to the Layer-by-Layer electrostatic self-assembly of polyelectrolyte multilayers

Eduardo Guzmán,[1,2,*] Ramón G. Rubio[1,2] Francisco Ortega,[1,2,*]

[1] *Departamento de Química Física I, Universidad Complutense de Madrid, Ciudad Universitaria s/n. 28040 Madrid, Spain*

[2] *Instituto Pluridisciplinar, Universidad Complutense de Madrid, Paseo Juan XXIII 1, 28040 Madrid, Spain*

[*] To whom correspondence should be addressed: eduardogs@quim.ucm.es (EG) and fortega@quim.ucm.es (FO)

**Abstract**

The fabrication of polyelectrolyte multilayer films (PEMs) using the Layer-by-Layer (LbL) method is one of the most versatile approaches for manufacturing functional surfaces. This is the result of the possibility to control the assembly process of the LbL films almost at will, by changing the nature of the assembled materials (building blocks), the assembly conditions (pH, ionic strength, temperature, etc.) or even by changing some other operational parameters which may impact in the structure and physico-chemical properties of the obtained multi-layered films. Therefore, the understanding of the impact of the above mentioned parameters on the assembly process of LbL materials plays a critical role in the potential use of the LbL method for the fabrication of new functional materials with technological interest. This review tries to provide a broad physico-chemical perspective to the study of the fabrication process of PEMs by the LbL method, which allows one to take advantage of the many possibilities offered for this approach on the fabrication of new functional nanomaterials.

# 1. The Layer-by-Layer method: from science to technology

The design of strategies for the modification of surfaces to protect them and avoid aesthetic effects has played an important role on the human technology from the antiquity to current days [1, 2]. This is exemplified in the current interest for the fabrication of coatings with controlled structure and specific functionalities by the modification of surfaces at molecular level, i.e. by a precise control of the structure of the components and their organization at the nanoscale level [3-6]. Functional nano-coatings allow tuning the properties and interactions of surfaces, which in turn are essential aspects on the development of different technological and industrial applications: fabrication of biocompatible films and drug delivery platforms, stabilization of colloidal solutions and particle dispersions, flocculation processes, surface treatments, cosmetics formulations, anticorrosion and antifouling surface and many others [7-11]. The possibility to control the surface properties and chemistry, almost at will, has been possible as result of the combined efforts of researchers belonging to different fields, including physics, chemistry, biology, medicine, engineering or material science [12-16].

The fabrication of functional materials with controlled properties in all the three dimensions remains a challenge yet. For much of the 20th century, the manufacturing of soft nano-nanomaterials was dominated by two types of assemblies [4]: the self-assembled monolayers (SAMs), and the Langmuir-Blodgett (LB) and Langmuir-Schaeffer (LS) films. The SAMs are ordered molecular assemblies adsorbed onto surfaces, with their formation being driven by a chemical specific interaction, e.g. thiols onto gold or silanes onto silicon surfaces [17, 18]. On the other side, the building of LB and LS films occurs by transferring amphiphilic molecules from a water/vapor intertace onto a solid surface, following a process which enables the deposition of multiple layers. However, these methods are limited to the fabrication of coatings onto flat macroscopic substrates, with a successful film deposition requiring the use of very clean substrates and dust-free atmosphere. Furthermore, LB and LS films present a limited mechanical and thermal resistance, and need long times for its fabrication [19-22]. The limitations of the SAMs, and LB and LS films were partially solved with the introduction of the Layer-by-Layer (LbL) method [11, 23].

The LbL method firstly introduced by Iler in the late 60's of the twentieth century for the electrostatic self-assembly of multilayers formed by the alternate deposition of oppositely charged colloidal particles [24] and revisited by Decher et al. [23, 25-28] almost three decades later, is nowadays a mature field which impact in different emerging areas of science and technology. This is mainly the result of the low cost, modularity, versatility and simplicity of this approach on the fabrication of multi-layered materials with tailored structure, well-defined thickness and composition and multiple functionalities, enabling the fabrication of materials including specific chemical, biological, optical, mechanical or electrical properties [9, 11, 29-31]. This has led to a rapid increase of the number of applications of materials obtained by the LbL in different fields of the nanotechnology (interfacial phenomena, colloids, and nanomaterials), including the fabrication of scaffold for tissue engineering, biomedical devices, wound healing dressing, encapsulation platforms, biosensors, cardiovascular devices, implants, conductive layers, perm-selective membranes, sensors, light-emitting thin films, sensors, electrochromic films, photonic systems, non-linear optical devices, antireflection coatings, self-healing and superhydrophobic surfaces [9, 31-44]. Additional indicators of the current interest of the LbL method are the more than 1000 papers related to the topic published each year, and the existence of some commercial products exploiting its advantages, e.g. contact lens coated by LbL films (Ciba Vision, Duluth, GE, USA) or coatings for chromatography column (Agilent Technologies, Santa Clara, CA, USA) [45].

The first works dealing with the fabrication of LbL films [25-27] took advantages of the well-known ability of polyelectrolytes to form self-organized supramolecular structures and their capacity to form complexes when are mixed with a polyelectrolytes bearing the opposite charge [46-48]. This allowed the fabrication of ultrathin polyelectrolyte films by the alternate deposition onto flat macroscopic substrates of layers of polyelectrolytes bearing opposite charges (or polyelectrolytes and bolaamphiphilies), the so-called polyelectrolyte multilayers (PEMs). However, the fabrication of LbL multilayers did not remain for long time limited to the electrostatic self-assembly (ESA) of synthetic oppositely charged polyelectrolytes and the list of compounds used for the fabrication of LbL films has been extended to other types of compounds (building blocks), both charged and uncharged, and currently the list includes colloidal particles and nano-objects (graphene and graphene oxide nanoplatelets, carbon nanotubes, dendrimers, clays, microgels, polymeric, ceramic or metallic particles), biomolecules (proteins and peptides, polysaccharides, nucleic acid, lipids), dyes, viruses, synthetic polymers and even in some cases small molecules [33, 38, 49-72].

The broad range of compounds that can be assembled using the LbL approach has made it possible the assembly of films using interactions of different nature than the electrostatic one, and there are many examples of the fabrication of LbL films through hydrogen bonding [73, 74], charge transfer interactions [75], molecular recognition [76, 77], coordination interactions [78], chiral recognition [79], host-guest interactions [80], $\pi$-$\pi$ interactions [81], biospecific interactions [82], sol-gel reactions [83], or even covalent bond ("click chemistry" reactions) [84, 85]. The only requisite to use an interaction in the fabrication of LbL films is that such interaction must be strong enough to ensure the chemical and mechanical integrity of material when they are exposed to stresses.

It has been stated that the LbL method was designed for the deposition of multi-layered coatings onto flat which means that the LbL approach must considered as a template-assisted assembly methodology. Therefore, the fabrication of LbL films requires a precursor substrate, sacrificial or not, that supports the assembly of the layers. Nowadays, the fabrication of LbL materials is possible using any type of solvent accessible surface as template, independently of its chemical nature, shape, geometry or size. Thus, the fabrication of LbL coating onto substrates which differ from the traditionally used macroscopic solid charged substrates is common, with colloidal micro- and nanoparticles, liposomes or vesicles, micelles, fluid interfaces (floating multilayers), emulsion droplets or even cells being some of the currently used templates [86-94]. The substrate may play different roles in the LbL assembly: (i) provides the geometry and morphology to the assembled material, remaining as a part of the final material once the deposition of the film is finished, or (ii) is only used as a template during the assembly process, and afterwards it can removed from the final material through chemical or physical procedures, allowing the fabrication of free-standing LbL materials, e.g. hollow capsules [95, 96] The versatility of the LbL method provides the bases for a careful control on the size, shape and morphology of the assembled materials, which allows the fabrication of different type of materials, from flat films to nano- and micro-capsules [29, 92, 97], and even multicapsules including several hierarchically organized nano-containers [98-102]. Furthermore, sophisticated materials including particles with onion-like structures, sponges, membranes or nanotubes has also manufactured taking advantage of the versatility and modularity of the LbL method [40, 103, 104]. Figure 1 presents some of the advantages and characteristics that made of the LbL method a key enabling technology on the fabrication of supramolecular materials with defined architecture.

Most of the studies on LbL materials, both theoretical and experimental, have paid attention to the growth mechanisms of the films, i.e. the dependence of the adsorbed amount of material on the number of adsorbed bilayers, N, and their potential applications [11]. However, there is an important lack of knowledge about certain aspects related to the physico-chemical bases underlying the formation of the LbL films and their properties. This is especially important when the internal structure and molecular properties of LbL multilayers, and their correlations are concerned [105-112]. The understanding of such aspects may only be obtained through a careful examination of the different variables that allows tuning the fabrication of LbL materials [113]: charge density of the assembled compounds and of the substrates [49, 114, 115], concentration [116], polyelectrolyte molecular weights [117], ionic strength [107, 111], solvent quality for the building blocks [118], pH [51, 119] and temperature [120]. The understanding of the impact of such parameters on the structure of the multilayers plays an essential role on the fabrication of supramolecular materials with controlled and tunable structures and physico-chemical properties. Therefore, the fabrication of materials with technological application making use of the LbL approach is a multidisciplinary challenge needing the study of the physico-chemical bases underlying the fabrication process. This review discusses the current understanding of the most relevant physico-chemical aspects for the application of LbL polyelectrolyte films in different technological and industrial fields. The extension of the research on LbL films makes it difficult to present a detailed discussion of the topic. Therefore, this work is focused on the study of those systems where electrostatic interactions play an important role on the assembly process, i.e. the assembly of polyelectrolyte multilayers. It is expected that the discussion contained in this review can help to close the gap between the physico-chemical knowledge on LbL polyelectrolyte multilayers and functional parameters with impact on the development of advances applications of this type of materials.

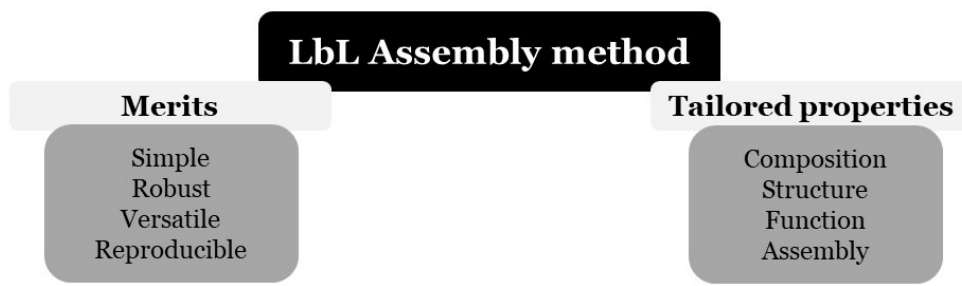

Figure 1. Advantages and characteristics of the LbL approach for the fabrication of functional materials.

## 2. Fabrication methods of Layer-by-Layer films

The fabrication of LbL films requires to establish assembly protocols taking into consideration that polyelectrolyte multilayers are thermodynamically unstable systems with respect to the formation of the respective inter-polyelectrolyte complexes in solution [113]. This is because the erosion of polymer chains from the multilayers is associated with a gain of the translational and configurational entropy of the chains, i.e. the mobility of the chains in the solution is higher than that of adsorbed molecules in the multilayer [16]. Furthermore the design of LbL materials with true technological applications requires the assembly methodologies easily scalable from the laboratory to an industrial level, which makes it necessary to engineer the LbL materials following the same rules used for the fabrication of bulk materials [121].

The LbL method is a simple and inexpensive procedure for the fabrication of multi-layered structures [58, 122]. Most of the LbL materials continue being assembled following procedures that are reminiscent from that introduced by the seminal works by Decher el al. [28]. This involves the deposition of the multilayers upon dipping the substrates into solutions of the components to be assembled. However, the adaption of the assembly process to the specific nature of the used templates (morphology, size or chemical nature) has required of slight modifications on the fabrication to manufacture an extraordinary variety of novel thin film materials. It is worth mentioning that the methodology used for the assembly influences strongly the properties of the obtained films, and their choice is depending on the final application of the material [123, 124].

Figure 2 shows a scheme representing the dipping deposition method for the deposition of LbL films onto flat macroscopic substrates. The deposition by dipping relies on the alternate immersion of the substrate into solutions containing the interacting compounds (polyelectrolyte bearing charges with the opposite sign or other type molecules), with intermediate rinsing steps between two consecutive adsorption steps [125]. This rinsing steps ensure that the material weakly adsorbed to the layer may be removed which avoid the cross-contamination during the assembly process. This is especially important when the assembly of polyelectrolyte is concerned because the formation of inter-polyelectrolyte complexes can result on their undesirable precipitation onto the LbL films, which may impact on the composition, structure and properties of the assembled material [45, 125]. It is worth mentioning that some attempts have been done to avoid the intermediate rinsing steps during the assembly of the LbL films, with dewetting method being a promising alternative for such purpose [126, 127]. The dewetting method involves the doping of the solution of the adsorbing compound with an organic solvent (dimethylformamide or dimethylsulfoxide), with the evaporation of such organic solvent and the subsequent dewetting process of the polymeric layer playing an essential role of the control of the layer deposition [126]. This makes that the time required for the assembly of the material can be reduced almost 30 times [127]. Another alternative on the seeking of faster deposition methodologies is to include a constant stirring, using a magnetic bar, of the solutions during the immersion of the substrates into the solution which allows limiting the deposition times to 10-20 seconds [128].

The fabrication of LbL films using the alternated dipping is rather slow because each adsorption and rinsing step requires several minutes (around 15 minutes per each deposition step) to ensure the diffusion of the molecules to the surface, their adsorption and the equilibration of the deposited films, which is a limitation for the industrial application of dipping deposition [44, 45, 129-131]. This has led to the implementation of different automatic dipping devices, which makes it easy the deposition process, even though the time required for the assembly of the materials remains long. However, the use of automatic dippers allowing the rotation of the substrate during its immersion in the solution provides the bases for an increase of 3-10 times of the deposition velocity, with the rotation velocity enabling the control of the layer thicknesses (thicker films are obtained by the increase of the rotation velocity).

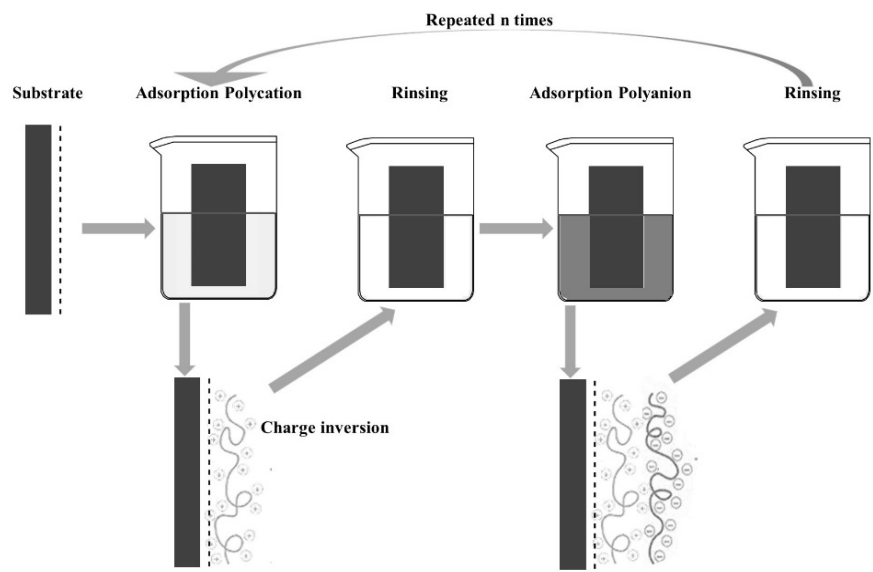

Figure 2. Scheme representing the different steps of the fabrication of LbL films by dipping.

The operational simplicity of the LbL approach has contributed to its adaptation to a broad range of situations. The spin coating and spray-assisted deposition are two methodological approaches, allowing for a speeding up the deposition process of LbL films [129, 132-135]. These approaches allows overcoming some of the main limitations of the assembly by dipping (time required for the fabrication of LbL films and difficulties for scaling-up), without compromising the properties of the obtained film properties. This is key on the development of LbL nanomaterials at the industrial scale. Figure 3 shows schemes of the spin coating and the spray-assisted deposition approaches.

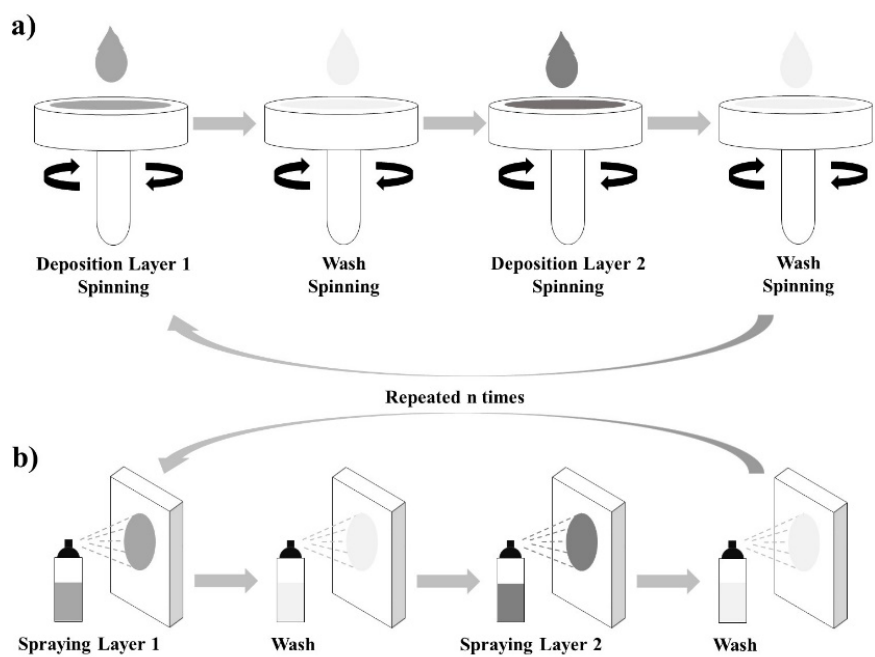

Figure 3. Scheme of the deposition of LbL films using spin-coating (a) and spray-assisted deposition (b).

The use of spin coating on the fabrication of thin films relies on the casting of the solution containing the compound to be assembled onto the substrate surface. Afterwards, the substrate undergo spinning at a constant velocity until the spreading of homogenous film onto the substrate and the complete evaporation of the solvent are attained. Once the layer is deposited, this is rinsed following a procedure analogous to that applied during the layer fabrication, with the substrate remaining under spinning until the total drying of the film is ensured. The repetition of the deposition and rinsing spinning cycles allows manufacturing multi-layered films [136, 137]. The use of spin coating deposition provides the bases for obtaining films with lower roughness and better organization than those obtained using the dipping approach, and a low degree of interpenetration between adjacent layers. This is explained considering the complex interplay between electrostatic and hydrodynamic interactions (centrifugal, air shear, and viscous forces) involved in the spinning-assisted deposition, with the former leading to the adsorption and rearrangement of polymers, and the latter governing the desorption of the weakly bound material from the surface and the dehydration of the films [137]. The intricate balance between electrostatic and hydrodynamic interactions helps to explain that spin coating can be several order of magnitude faster than dipping and the smaller thickness of the obtained films, with the latter being related to the spinning velocity (higher velocities lead to thinner films). Furthermore, spin-coating deposition leads to a significant reduction the amount of material required for the deposition of the layers [137]. However, the advantages of spin-coating cannot overcome the important drawbacks associated with the low volatility of the water which is the solvent commonly used for LbL assembly, limiting the practical application of spin-coating on LbL deposition. It is worth noting that the success of the deposition of layers using spin-coating requires a careful control of the spinning velocity and solution concentration, with both parameters being essential affecting to the thickness of the deposited layers [138].

An alternative way for exploiting the rotation of the substrate on the fabrication of LbL films is the deposition under high-gravity fields, using rotating media, introduced by Ma et al. [139] for the fabrication of multilayers formed by poly(ethylenimine) (PEI) and zirconia nanoparticles. The main advantage of this deposition methodology is associated with the hastening of the diffusion process, which shortens the time required for reaching the adsorption equilibrium. This methodology provides the bases for obtaining LbL films with similar quality than those obtained by dipping, with a less time consumption, which is rationalized considering the increase in the concentration gradient and the importance of the turbulence in the deposition process.

The use of the spray-assisted deposition for the fabrication of LbL materials was firstly introduced by Schlenoff et al. [140] for the alternate deposition of poly(4-styrene sulfonate of sodium) (PSS) and poly(diallyl-dimethylammonium chloride) (PDADMAC), and relies on the deposition of layers by an alternate spraying of solutions containing different materials onto the surface of a flat substrate, with intermediate rising cycling between the deposition of two adjacent layers. It is worth noting that most of the advantages stated for the spin-coating assembly also apply when the spray-assisted deposition is concerned [140], with the roughness and the thickness of the multilayers obtained by spraying being significantly smaller than those of films obtained by dipping [124, 141-143]. This may be understood considering the coexistence of two simultaneous process during the spraying of the solutions: (i) adsorption and (ii) drainage. This requires performing the spraying perpendicularly to the surface to ensure the gravitational drainage which favours a fast removal of the excess of sprayed solution. However, the gravitational drainage may lead the inhomogeneous films, with an enhanced deposition close to the solution drips [134, 144]. This may be partially solved by using rotating substrate during the deposition process [145-147]. The spray-assisted deposition allows the reduction of the contact time (less than 10 second may be enough) between the adsorbing material and the surface which results on a limited interpenetration between adjacent layers [134]. This has led to a strong development of the application of spray-assisted deposition at industrial level [44, 148]. A further advantage of the spray-assisted deposition of LbL films is that this methodological approach enables a fast fabrication of uniform LbL films on substrates with a large area [4, 149]. The deposition by spraying can be furtherly sped-up using vacuum which allows minimizing the lag time between the different steps involved in the fabrication process [44]. The effectiveness of the spray-assisted deposition requires the optimization of the spray-substrate distance and the times of spraying and draining [44]. Furthermore, the solution concentrations, the volume and flow of the sprayed solutions, the spraying time and the waiting time between two consecutive steps and whether rinsing steps are included or not are essential and how long these rinsing steps are important aspects to considered when the fabrication of materials using the spray-assisted deposition is concerned [134, 140, 143-145, 150-152]. Table 1 summarized some differences on the characteristic of the films obtained using the traditionally dipping deposition and spin-coating or spray-assisted deposition.

Table 1. Characteristic of LbL films obtained using dipping deposition and spin-coating or spray-assisted deposition.

| Dipping | Spin-coating/Spray-assisted deposition |
|---|---|
| Thick | Thin |
| Rough | Smooth |
| Interpenetrated | Stratified |
| Opaque | Transparent |

An alternative to the conventional spray-assisted LbL deposition consists in the simultaneous spraying of two or more interacting species against a substrate, the so-called simultaneous spray coating of interacting species (SSCIS). This type of spray-assisted deposition leads to a fast interaction between the complementary species on the surface, enabling a continuous and gradual growth of the films. The thickness of the obtained layers depends on the spraying time, with the solvent and excess of material being removed by drainage [151, 153]. It is worth mentioning that the thicknesses of the multilayers fabricated by the SSCIS approach were found to be similar to those of films obtained using the conventional alternate spray-assisted approach [153].

The use of electric or magnetic fields can help on the deposition of LbL films [154, 155]. This leads to denser and thicker films than any other technique used on the assembly of LbL materials [156, 157]. The application of electric field is generally based in the use of electrodes as templates for the assembly of the films (electrodeposition). This allows the deposition of the films by the application of a voltage in electrolytic cells. The standard electrodeposition approach for the fabrication of LbL films requires the immersion of two electrodes in the solution of the polymer to be deposit, followed by the application of an electric current within the deposition cell to drive the formation of the layer. Once the layer is deposited, the electrodes are rinsed with the solvent, and then they are placed in a solution containing the compound forming the second layer and the process is repeated [158]. Several experimental designs has been used for optimizing the electrodeposition of LbL films, which can be applied to the deposition of multi-layered structures onto macroscopic surfaces or colloidal particles [127]. The control of the thickness and roughness of the electrodeposited films can be done optimizing the applied voltage and the time of the process [159].

The fabrication of electrodeposited films can be performed taking advantage of local effects appearing at the electrodes surfaces, e.g. redox reactions or pH changes. Therefore, the presence of electrodes may induce a local pH change in their vicinity in relation to that of the solution, with the existence of a lower value of pH close to the anode which allows the deposition [160]. However, this approach is limited to the deposition of a few layer because the formation of a layer reduce the penetration of the current towards the electrode, with the disappearance of and the effect of the pH change.

Magnetic fields can be also exploited for the deposition of LbL films [161]. This approach is based in the immersion of the substrates in solutions containing charged magnetic sensitive materials, which are deposited following the traditional LbL dipping approach. Magnetic fields are applied between the deposition steps of two adjacent layers to control the packing of the layers, and consequently the multilayer thickness [162].

The above discussion was devoted so far of the fabrication of LbL films onto macroscopic surfaces. However, many applications of LbL materials are associated with the encapsulation of active compounds [9, 38, 104]. This makes it necessary to design deposition methodologies enabling the use of colloidal particles, micelles, liposomes, vesicles or emulsion drops as templates. The use of colloidal templates in LbL deposition is based in the deposition by immersion (dipping) firstly introduced by Decher et al. [28]. However, some modifications must be introduced on the deposition procedure accounting for the specific characteristics of the colloidal substrates [86, 87, 163].

One of the most important aspects to consider, when colloidal substrates are used as template for LbL deposition, is that these are commonly dissolved or suspended in a solvent, generally water.

The seminal works dealing with the LbL coating of colloidal particles were done by the group of Möhwald in the late 90's of the past century [86, 87, 164]. Their approach was based on the alternate deposition of polyelectrolytes bearing opposite charges onto colloidal microparticles which intermediate washing cycles. This allows the fabrication of both core-shell structures and hollow capsule. The template on the latter cases is a sacrificial substrate which can be removed in a final step, generally by a chemical dissolution process. This depends on the chemical nature of the colloidal template, e.g. fluoride acid is used for removing $SiO_2$ particles, diluted hydrochloride acid for melamine formaldehyde resins and tetrahydrofuran when polystyrene latex particles are used as templates [11]. One of the most important challenges in the preparation of LbL materials using colloidal templates is the separation of the excess of non-adsorbed material from the dispersion containing the particles capped with the multilayer. This is affordable when microparticles are used as templates by the centrifugation of the suspension after the adsorption step. The centrifugations leads to the sedimentation of the decorated particles sediment at the bottom which allows removing the supernatant, where the excess of non-adsorbed material remains. Once the supernatant is removed, the particles are re-dispersed in an aqueous medium, and the centrifugation-re-dispersion steps are repeated, commonly three times, to ensure that the material excess is completely removed. Once the clean suspension of particles decorated with the multi-layered structure is obtained, the adsorption of the next layer is performed repeating the sequence of adsorption and cleaning. This methodology minimizes the production of insoluble inter-polyelectrolyte complexes in the aqueous phase during the assembly process [165, 166]. Figure 4 shows a sketch of the different steps involved on the deposition of polyelectrolyte multilayers onto a negatively charged particle.

The application of the above described methodology is difficult when nano-sized colloids are chosen as templates. This is because their sedimentation using centrifugation is not easy and the application of alternative procedures for the separation of the particles decorated with the multilayers and the excess of non-adsorbed material are required [167]. One of the most promising alternatives for such purpose is the serum replacement method, which also helps on the preparation of highly concentrated capsule suspensions [168]. It is worth mentioning that the use of filtration processes, e.g. serum replacement, for the separation of the capsules and the excess of non-adsorbed material provides the bases for increase the velocity of the LbL assembly, enhancing the recovery yield. The latter is essential for an industrial scaling up of the assembly process, with the preparation of concentrated capsules suspensions remaining as an important challenge for the LbL method. This has been partially resolved by using tubular flow type reactors, which allows preparing capsules with the required number of layers following a procedure of production in continuous. However, a limitation for such method is that a reduced amount of the polyelectrolyte forming the last deposited layers always remains in solution after each deposition step, which can lead to cross-contamination as result of the formation of inter-polyelectrolye complexes [169].

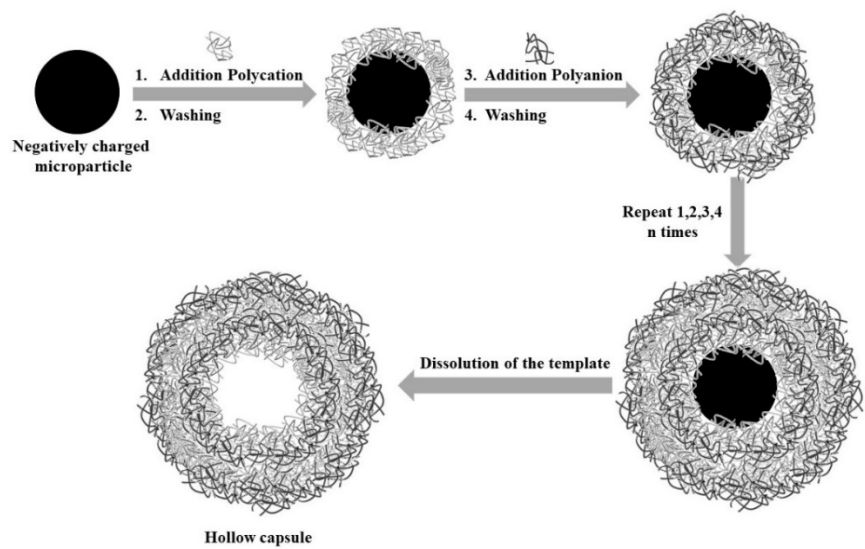

Figure 4. Scheme representing the methodology used for the fabrication of LbL materials using as template a negatively charged colloidal microparticle. The representation shows the layers assembly steps, and the final treatment for removing the template to obtain a hollow capsule.

The above described approaches are generally useful when particles with higher density than the water are used. However, the use of particles lighter than water (e.g. emulsion drops, vesicles or liposomes) needs to adapt the methodology, and in particular the separation methodologies. In the particular case of the deposition of LbL films onto vesicles or liposomes, the procedure can be summarized in the following steps [170]: (i) a solution containing the material to be assembled in the first layer is added to a diluted suspension containing the vesicles or liposomes; (ii) once the first layer is deposited, a solution containing the material used for the second layer is added to the dispersion of coated vesicles or liposomes and the deposition of the layer occurs, whereas the excess of polyelectrolyte leads to the formation of inter-polyelectrolyte complexes with the non-adsorbed polyelectrolyte remaining in the solution after the first adsorption step. This inter-polyelectrolyte complexes are depleted as a precipitate form the solution; (iii) centrifugation of the dispersion to enhance the separation of the precipitated inter-polyelectrolyte complexes and the supernatant containing the capsules. The application of the centrifugation is possible because the formation of the polyelectrolyte shell provides enough rigidity to the vesicles/liposomes, avoiding their aggregation and fusion; and (iv) after the deposition of the first bilayer, the process can be repeated several times up to the required number of layers is obtained. However, the formation of inter-polyelectrolyte, which may form supramolecular aggregates with the decorated liposomes, and the centrifugation process leads to a reduction of around 5% of the total number of liposomes/vesicles after each adsorption/separation cycle, which limits the number of layers to values no higher than 10-12 [170]. It is worth mentioning that the use of emulsion drops, in particular of oil in water emulsions, as templates for the deposition of LbL films has received a big attention in recent years. This is because emulsion drops can be used as reservoir where an active molecules can be solubilized, enhancing its bioavailability [171-174].

One of the most important challenge of the fabrication of LbL films onto light-weight colloidal templates is to avoid the centrifugation steps because it can lead to the aggregation of the templates and/or the formed capsules. A possible alternative for accomplishing such purpose is to remove the cleaning step from the fabrication procedure by the addition of the exact amount of polyelectrolyte required for reaching the surface saturation [86, 175]. The saturation method allows the fabrication of supramolecular structures to those obtained by the introduction of centrifugation steps, with an increase of the velocity of the assembly process by a factor 3 [86, 174]. The optimization of the saturation procedure requires a careful examination of the zeta potential values during the deposition procedure [86].

Another alternative for avoiding the centrifugation is to immobilize the particles in an agarose hydrogel. This allows one to consider the particles collected within the immobilization matrix as a planar substrate where the deposition can be performed using the traditional dipping approach. Once the desired number of layer are deposited, the agarose hydrogel can be removed by a heating procedure (at 37ºC) followed by a cleaning procedure based on the repetition of centrifugation-re-dispersion cycles at least three times. The use of this type of deposition results on the formation of thinner films than those obtained using the conventional approach used for coating colloidal particles, which may be explained in terms of a limited diffusion of the polymer through the agarose hydrogel [176].

In recent years, the use of microfluidic systems for the fabrication of LbL films is gaining interest. This type of devices enables the deposition of LbL layers onto both the channel walls and substrates placed or immobilized within the channels [177-180]. The most common approach for taking advantage of the microfluidic on the fabrication of LbL films involves the use of pressure or vacuum to drive the sequential displacement of the polymer and washing solutions within the microfluidic chips, with the time of contact being the most important factor governing the deposited amounts [181-183]. The interesting perspectives that microfluidic opens in LbL deposition are limited by the difficulties associated with the optimization of the assembly process in each particular case and the high price of the instrumentation [127].

The above discussion shows clearly that LbL deposition can be performed independently on the chemical composition or surface charge of the substrate. Furthermore, it is worth mentioning that, in most of the cases, the thickness of the layers appears to be rather independent on the substrate, thus after the deposition of a certain number of layers, i.e. once a certain degree of coverage has been reached, the deposition of additional layers is not significantly affected by the surface nature [45]. Furthermore, a detailed understanding of the LbL assembly of polyelectrolyte multilayers makes it necessary to explore the physico-chemical aspects influencing the formation process and the final properties of the films, with two aspects being essential for such understanding: (i) polyelectrolyte multilayers are highly hydrated systems and contain counterions, and (ii) the adsorption of polyelectrolyte layers is almost irreversible.

## 3. Physico-chemical aspects of the electrostically-driven Layer-by-Layer method

### 3.1. Growth mechanisms of Layer-by-Layer multilayers

One of the most important aspect of the LbL method is the possibility of tuning the thickness of the films almost at will. This requires a careful examination of the dependences of the adsorbed amount (or layer thickness) on the number of bilayers, N, i.e. the growth mechanism. Two different dependences are frequently found for the growth of PEMS: linear and non-linear, with the latter being generally referred in the literature as exponential growth, even though the dependence of the adsorbed amount on the number of bilayer is not truly exponential (Notice that the films are denoted as $(A - B)_n$, with A and B representing the interacting species (commonly polycation and polyanion, respectively) and the subindex n representing the number of bilayers). Figure 5 presents a scheme of the two common growth dependences appearing in LbL films.

The existence of linear growth is characterized by a quasi-linear dependence of the adsorbed amount on the number of bilayers, i.e. the adsorbed amount increases quasi-linearly upon the adsorption of a bilayers. In most of the case, polyelectrolyte multilayers presenting a linear growth evidence an increase of the thickness around a few nanometers after the adsorption of a bilayer, which is roughly the sum of the characteristic lengths of the polycation and the polyanion. This thickness can be modified by playing with the conformation of the polyelectrolytes in the solution, which is possible by the modification of the assembly conditions, i.e. the charge density for weak polyelectrolytes or the balance of interactions occurring in the solution. The linear growth is generally found in (PAH - PSS)$_n$ multilayers (with PAH being poly(allylamine hydrochloride)) almost independently on the assembly conditions and for (PDADMAC -PSS)$_n$ films assembled under conditions in which a high effective charge density exists in both polyelectrolytes (commonly assembled from low ionic strength solutions) [184, 185]. Other examples of multilayers growing in a linear way are (PAH - PAA)$_n$ (PAA being and (PM2VP - PSS)$_n$ (with PAA and PM2VP being poly(acrylic acid) and poly(N-methyl-2-vinyl pyridinium chloride), respectively) [186] The non-linear growth is characterized by an increase of the adsorbed amount faster than that found in a linear growth (supra-linear dependence). Among the systems presenting non-linear growth are accounted multilayers combining PDADMAC and PSS assembled under conditions in which a reduced effective charge exists in the polyelectrolytes (generally LbL films assembled from solutions with high ionic strength) [107, 111, 118, 187], and different multilayers including biopolymers among their components, e.g. (CHI – PAA)$_n$, (PLL - HA)$_n$ or (PLL - PGA)$_n$ –with CHI, PLL, HA and PGA being chitosan, poly(L-lysine), hyaluronic acid and poly(glutamic acid), respectively- [50, 51, 188-190]. Table 1 reports some examples of multilayers growing linearly and non-linearly.

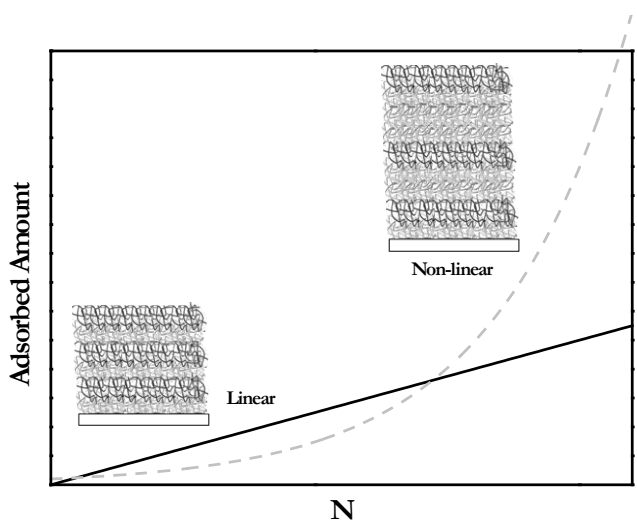

Figure 5. Scheme representing an idealization of the two common growth dependences appearing in LbL films: linear and non-linear.

Table 2. Summary of some multilayers presenting linear growth and non-linear growth.

| System | Ref. | Notes |
| --- | --- | --- |
| *Linear Growth PEMs* | | |
| (PDMAEMA-PSS)$_n$ | [49] | |
| (PDADMAC-PSS)$_n$ | [89, 107, 111, 118, 184, 191] | PDADMAC with a high effective charge density (low ionic strength conditions) |
| (PAA – PLL)$_n$ | [117] | Assembly using polyelectrolytes with low molecular weight |
| (PAH - PAA)$_n$ | [119, 186] | Assembly dependent on the pH and charge density of the polymers |
| (PAH-PSS)$_n$ | [184, 185, 191] | Almost independent on the conditions used for the assembly |
| (PM2VP - PSS)$_n$ | [186] | |
| (PDMA - PMAA)$_n$ | [192] | |
| (PEI - DNA)$_n$ | [193] | |
| (PEI - BSA)$_n$ | [194] | |
| (PEI - PAA)$_n$ | [195] | |
| (CHI – CMC)$_n$ | [196] | |
| (CHI - ALG)$_n$ | [197, 198] | |
| (PEI - ALG)$_n$ | [197] | |
| *Non-Linear Growth PEMs* | | |
| (CHI - PAA)$_n$ | [50, 51, 199] | |
| (PDADMAC-PSS)$_n$ | [89, 107, 111, 118, 184, 191] | PDADMAC with a low effective charge density (low ionic strength conditions) |
| (PAH - PAA)$_n$ | [119] | Assembly dependent on the pH and charge density of the polymers |
| (PAH-PSS)$_n$ | [120] | Assembly at high temperatures (T > 55 ° C) |
| (PLL - HA)$_n$ | [188, 189, 200, 201] | |
| (PAH - HA)$_n$ | [199] | |
| (CHI - HA)$_n$ | [199, 202] | |
| (PLL - PGA)$_n$ | [203] | |
| (CHI - PGA)$_n$ | [204] | |
| (CHI - FUC)$_n$ | [205] | |
| (PLL - HEP)$_n$ | [206] | |
| (CHI - HEP)$_n$ | [207] | |

[1] PDMAEMA: triblock copolymer PDMAEMA-PCL-PDMAEMA, with PDMAEMA and PCL being poly[2-(N,N-dimethylamino)ethyl methacrylate) and poly(epsilon-caprolactone), respectively; PSS: poly(4-styrenesulfonate of sodium); PDADMAC: poly(diallyl-dimethylammonium chloride); PAA: poly(acrylic acid); PLL: poly(L-lysine); PAH: poly(allylamine hydrochloride); PM2VP: poly(N-methyl-2-vinyl pyridinium chloride); CHI: chitosan; HA: hyaluronic acid; PDMA: poly(2-(Dimethylamino)ethyl methacrylate); PMAA: poly(methacrylic acid); PEI: poly(ethylenimine); DNA: Desoxyribonucleic acid; BSA: bovine serum albumine; CMC: carboxy(methyl cellulose); ALG: alginate; PGA: poly(glutamic acid); FUC: fucoidan; HEP: heparin

The differences in the dependence of the adsorbed amount on the number of bilayers appearing in polyelectrolyte multilayers have been justified in terms of different physical mechanisms. The first attempt for providing a physical picture accounting for such differences was performed by Elbert et al. [7]. They consider the existence of diffusion of the polyelectrolytes within the multilayer structure in both linear and non-linear growth multilayers. However, such mobility of the chains presents subtle differences depending on the nature of the polyelectrolytes, and the type of growth. The study by Elbert et al. [7] assumed that the formation of linear growth multilayers occurs by an initial deposition of molecular polyelectrolyte layers, which can be strongly intermingled due to the diffusion of the polyelectrolyte chains within the perpendicular direction to the film surface, resulting in a blend layer. The final thickness of such films is determined by the molecular weights and the charge characteristics of the polyelectrolytes. The situation was found to be slightly more complex when multilayers presenting a non-linear growth were analysed. This type of growth is assumed to be driven for the deposition of more than one layer in each deposition cycle, which should be limited as result of the strong repulsion between polyelectrolytes chains. However, this may occur considering that the diffusion of polyelectrolytes within the multilayers structure may results in some cases (specific polyelectrolyte mixtures and assembly conditions) in a coacervation process between the polyelectrolyte in solution and the oppositely charged polyelectrolyte chains diffusing from the multilayer. These coacervates can precipitate onto the multilayer surface resulting in an increase of the thickness higher to than expected for a molecular layer [208, 209] in agreement with the theoretical studies by Tang and Besseling [210]. These latter author found a correlation between the behaviour of inter-polyelectrolyte complexes in solution and that corresponding to polyelectrolyte multilayers. Furthermore, the existence of different dynamics of the polyelectrolytes in linear and non-linear growth multilayers are in agreement with the finding by Xu et al. [211]. They attributed the transition from linear growth films to non-linear growth one in (PDMA-PMAA)$_n$ multilayers to a change on the dynamics of the polyelectrolytes changed occurring as result of a modification of the ionic pairing within the multilayer.

An alternative explanation that rule out the role of the diffusion in linear growth polyelectrolyte multilayers was proposed by Picart et al. [188, 189, 200, 201, 212]. They explained that non-linear growth for (PLL – HA)$_n$ multilayers results from an *in* and *out* diffusion of one of the polyelectrolytes within the multilayer structure. In particular, this diffusion of the PLL chains diffusion within the multilayer structure was found, with this diffusion being depending on the nature of the solution interacting with the multilayer. Thus, the exposure of (PLL – HA)$_n$ to PLL solutions leads to a diffusion of the polyelectrolyte to the inner of the multilayer, whereas the opposite in true when the multilayers are exposed to HA solutions. The diffusion of the PLL to the solution-film interface drives the PLL-HA complexation, favouring the adsorption of more HA than that corresponding to a single monolayer, with the overall growth resembling a situation in which each bilayer is thicker than those expected from the molecular sizes of the polyelectrolytes, resulting in a non-linear growth of the film [188]. The diffusion of polyelectrolytes on (PLL - HA)$_n$ films was furtherly analysed by Picart et al. [212] using confocal laser scanning microscopy (CLSM) to monitor the adsorption process within the axis perpendicular to the multilayer by adding fluorescently labelled polyelectrolytes at different height of the multilayer. The results showed that whereas the fluorescently labelled PLL was able to diffuse within the entire multilayer, the fluorescently labelled HA remained at the fixed positions where it were assembled. This was claimed by Picart et al. [212] and by Lavalle et al. [213] as a clear evidence of the role of the interdiffusion of at least one of the polyelectrolyte, and in some cases both of them -(PLL-PGA)$_n$ films-, as driving force of the non-linear growth [182]. However, the works by Picart et al. [188, 212] were only focused on films with non-linear growth, and no control experiments on multilayers presenting linear growth were included. Thus, even the driving force of the non-linear growth proposed by Picart et al. [188] agrees qualitatively with the picture provided by Elbert et al. [7], i.e. diffusion of one of the polyelectrolytes combined with the coacervation and precipitation of inter-polyelectrolyte complexes onto the polyelectrolyte surface, Picart et al. [188, 212] rule out the existence of interdiffusion in multilayers presenting grow linearly. The inconsistencies of the model were revised by Porcel et al. [214]. They introduced the concept of the restructuring inner compartment which is a region of the multilayer where the polymer mobility is hindered due to its high density. The formation of this zone leads to a situation in which the concentration of polymer diffusing along the multilayer is constant irrespectively of the multilayer thickness. However, this refined model was only to describe the behaviour when the molecular weight of the polymers is relatively high (PLL with molecular weight around 360 kDa), with the diffusion within the entire multilayer being found for smaller molecular weights (20 kDa).

The interdiffusion of at least one of the polyelectrolyte within the multilayers was associated by Lavalle et al. [201] and Hoda and Larsson [215] with the existence of a Donnan effect along the LbL film. This Donnan effect results from the mobile charges generated by the *in* and *out* diffusion of the polyelectrolyte chains, which leads to an excess of charge within the multilayer, with the interdiffusion occurring until such charge excess is completely compensated. However, this framework does not provide a satisfactory explanation of the transition from linear to non-linear growth occurring in some multilayers due to changes of the conditions used for the assembly of the layers, e.g. (PDADMAC – PSS)$_n$, [107, 187, 216].

The existence of mobility of the polyelectrolyte chains within the whole multilayers proposed by Elbert et al. [7], independently of the growth mechanism of the multilayers, was demonstrated by Guzman et al. [184] from a rigorous analysis of the adsorption kinetics of the layers, which has evidenced that the interdiffusion is not limited to non-linear growth materials. This suggests that the differences in the multilayer roughness can contribute to the emergence of two different types of growth as was evidenced for (PDADMAC – PSS)$_n$ multilayers [187], and contradicts the conclusions obtained by Lavalle et al. [189]. However, the latter study neglects the impact of the specific chemical nature of the assembled polyelectrolytes, already discussed in the work by Elbert et al. [7], when the comparison between linear and non-linear growth multilayers is performed.

The impact of the roughness as driving force for the transition from a linear to a non-linear growth of the polyelectrolyte multilayers can be understood considering that the increase of the multilayer roughness leads to an increase of the area available for the adsorption. This results on an increase of the adsorbed amount in the successive deposition cycles, and as matter of fact on a non-linear growth. On the other side, the roughness remains almost constant with the increase of the deposition cycles when linear growth multilayers are concerned. The transition from linear to non-linear growth may be ascribed to the different conformation of the adsorbed polymer chains [184, 187, 191, 216, 217]. The impact of the roughness in the growth of polyelectrolyte multilayers is compatible with the results by Haynie et al. [218] in which the emergence of a non-linear growth is the result of the propagation, growth and coalescence of dendritic or isolated structures, which leads to the increase on the roughness of the films. The formation of heterogeneous structure during the growth leads to non-linear dependences of the thickness on the number of layers because the size of the deposited material results in the increase of the area available for the deposition. The transition between the region of heterogeneous growth to that where a homogenous growth is found may be explained as result of sterical limitations for the polyelectrolyte chains, leading to a constant number of binding sites [219]. This agrees with the results by Hernández-Montelongo et al. [202] using fractal analysis on Atomic Force Microscopy images. They found that the assembly of linear growth (CHI-HA)$_n$ multilayers follow a pathway based on the aggregation of the molecules followed by their rearrangement, whereas those multilayers presenting non-linear growth follow an assembly governed by a diffusion limited aggregation. This process results in the formation of aggregates due to diffusion gradients, with depleted zones remaining in the layer, which result in the formation of rough multilayers with irregular surfaces. On the contrary, linear growth multilayers present a smoother and regular surface. It is worth noting that the existence of region with heterogeneous growth was also observed by Picart et al. [188] during the initial stages of the deposition of (PLL-HA)$_n$ multilayers, even though they neglect its impact on the emergence of the non-linear growth of the multilayers. It is worth mentioning that the formation and growth of such heterogeneities within the polyelectrolyte multilayer do not limit the interdiffusion to non-linear multilayers.

The above described dependences of the adsorbed amount on the number of bilayers are the most commonly found in polyelectrolyte multilayers. However, it has been reported more exotic growth dependences on specific polyelectrolyte multilayers, e.g. multilayers of a short polyanion, poly(sodium phospate), and PAH. Such exotic supramolecular films follow a growth mechanism which violate most of the rules driving the assembly of LbL polyelectrolyte films [220, 221], e.g. the multilayers growth without the typical charge inversion expected in polyelectrolyte multilayers and it presents growth instabilities.

It is clear that the explanation of the possible origin of the two growth mechanisms in polyelectrolyte multilayers remains far from clear. However, different studies suggest that the interdiffusion of the polyelectrolyte chains within the multilayers cannot be considered the distinctive signature for the transition between linear to non-linear growth, with such interdiffusion appearing independently on the type of growth appearing on the multilayers. On the other side, the roughness model may be suitable to explain deviations from a linear growth only during the deposition of the first few layers. Thus, it may be assumed that the origin of the specific growth mechanism of polyelectrolyte multilayers can be only understood from a careful examination of the specific characteristics of the assembled polyelectrolytes and their conformations, the conditions used for the assembly of the multilayers and the interactions involved in the assembly process.

## 3.2. Charge compensation and charge overcompensation: driving the assembly of multilayers

The use of charged compounds in the formation of LbL makes it necessary to consider the important role of the electrostatic interactions between the polyelectrolyte adsorbed in adjacent layers as driving force of the assembly. However, the complete understanding of the assembly process requires a careful examination of the complex interplay between the different interactions involved: polyelectrolyte – polyelectrolyte, polyelectrolyte – solvent and polyelectrolyte – template [111, 222]. The understanding of these interactions makes it necessary to consider the impact of two different aspects on the assembly: (i) quality of the solvent for the polyelectrolytes (ionic strength, pH or temperature), and (ii) competence between electrostatic and entropic factor [187, 216].

It is commonly accepted that the deposition of charged polymers or particles onto oppositely charged surfaces is driven for a charge inversion phenomenon, i.e. the deposition of layer is not stopped when the neutralization of the charge of the surface is reached, with the adsorption proceeding until a certain degree of opposite charge appears on the surface in accordance to the Surface Force Measurements by Berndt et al. [223]. This phenomenon can be understood considering that the adsorption of a polyelectrolyte monolayer onto a surface bearing opposite charges does not results initially on the neutralization of the surface due to steric factors, and hence it is necessary the presence of additional charges chains to ensure the neutralization of the uncompensated charges of the substrate. This leads to the charge inversion and to the formation of layers with charged loops and tails protruding to the solution, which overcompensates the charge of the initial surface and hinders the adsorption of additional molecules as result of the repulsive electrostatic interaction (the LbL assembly of polyelectrolyte layers is an electrostically self-limited process). The immersion of the surface decorated with a polyelectrolyte layer into a solution containing a polyelectrolyte bearing the opposite charge allows the deposition of a new polyelectrolyte layer following a similar pathway to that described for the previous layer, with the charge reversion, i.e. the overcompensation of the charge of the previous layer enabling the alternate deposition of layers of building blocks with opposite charge [51, 88, 187, 224]. It is worth mentioning that even though the charge inversion phenomena may appear counterintuitive, it is essential for the electrostatic self-assembly of LbL films [111], and for many other processes with interest for materials science and biology [225].

The overcompensation during the deposition of LbL films has been commonly studied for multilayers deposited onto macroscopic substrates and onto colloidal templates in term of different experimental parameters which provide information related to the charge of the surface: surface potential (measured using a Kelvin probe), streaming potential or the zeta potential ($\zeta$ potential, obtained from electrophoretic mobility measurements) [51, 88, 107, 111, 175, 185, 187, 188, 226, 227]. This type of measurements shows the switching of the surface charge between positive and negative values with the alternate deposition of polycation and polyanion layers, respectively. Figure 6 shows the changes of the value of the $\zeta$ potential for (PAH-PSS)$_n$ multilayers deposited onto silica colloidal particles (1 μm). The impact of the overcompensation is clear with the $\zeta$-potential oscillating between values around (30 ± 10) mV, when the adsorption of polycation layers is concerned, and (-40 ± 10) mV after the deposition of PSS layers. This agrees with the assymetric growth of polyelectrolyte multilayers suggested by Ghostine et al. [228], which involves a different degree of charge overcompensation depending on the nature of the deposited layer. The overcompensation is maximum at the surface of the layers, and is expected to decay exponentially with the penetration toward the inner region of the multilayers [107, 111].

The degree of charge inversion occurring upon the deposition of polyelectrolyte layers is self-limiting by the specific pair of assembled polyelectrolytes, without any significant impact of the conditions used for the assembly (ionic strength, pH) on the maximum degree of charge inversion reached [187, 216]. However, the extension of the overcompensation within the layer thickness depends on the growth mechanism and the fuzziness of the obtained multilayer: (i) multilayers exhibiting non-linear growth present overcompensation throughout the entire layer [189], and (ii) multilayers exhibiting linear growth present overcompensation which is almost limited to the layer surface [23].

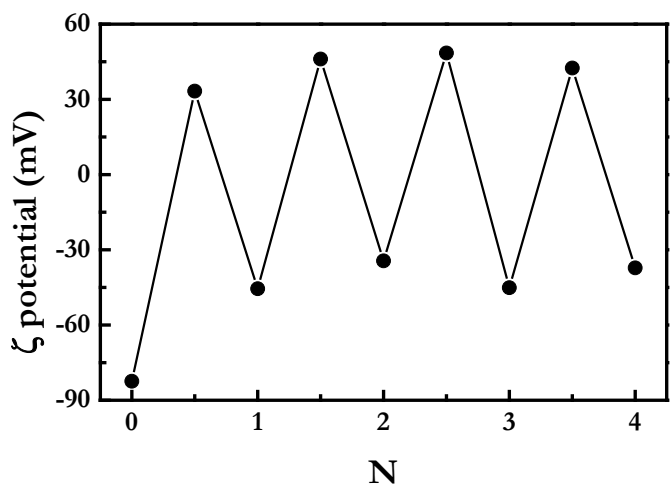

Figure 6. Change of the ξ potential with the alternate deposition of PAH and PSS layers onto silica microparticles (1 μm) from polyelectrolyte solutions with concentration 1 mg/mL and ionic strength fixed in 10 mM. Notice that N = 0 corresponds to the microparticles, N=1, 2, 3 and 4 correspond to the deposition of PSS layers, and N= 0.5, 1.5, 2.5 and 3.5 correspond to the deposition of PDADMAC layers.

The existence of overcompensation is correlated with an excess of polymer charges which are not paired with those in the adjacent layer. The existence of such charge excess makes the multilayer unstable, with its stability requiring the fulfillment of the electro-neutrality boundary condition [107, 229], i.e. it is needed to introduce a contribution enabling to counter-balance of the charge excess associated with the assembly of the polyelectrolyte layers and ensuring a zero net charge at the macroscopic scale (beyond the Debye length). This is only possible considering the incorporation of small ions which compensate the charge excess associated with unpaired segments of the polyelectrolyte chains [118, 187, 230]. The possible incorporation of such small ions makes it mandatory to examine the mechanism driving the charge compensation, the so-called compensation mechanisms. These are required because the equilibria between the different charges species appear coupled, i.e. the entrance of a cationic specie into the multilayer only can occur coupled to the entrance of an anion or the creation of a negative charge within the multilayer, and vice versa [231-233].

It is accepted the existence of two different mechanisms for the compensation of charge in polyelectrolyte multilayer: (i) intrinsic, and (ii) extrinsic [229]. The former mechanism (intrinsic) involves the absence of polymer charge excesses during the LbL assembly, i.e. the charges of the polyelectrolyte deposited in one layer are directly compensated by the charges of the polyelectrolyte adsorbed in the adjacent layer, which results in the formation of multilayers with stoichiometry 1:1 (polycation:polyanion). This results in the highest level of ionic cross-linking that can be expected for the assembly of polyelectrolyte multilayers. The formation of intrinsically compensated multilayers requires the release of a large amount of counterions (condensed counterions to the polyelectrolyte chains) from the polymeric film to the solution which decreases the average free energy of the system due to the important contribution of the entropy increase associated with the release of counterions. Thus, the entropy gain associated with the release of the counterions becomes the driving force of the assembly of intrinsically compensated LbL films. On the other side, those systems where an extrinsic compensation is found need of the presence of counterions to ensure the electroneutrality of the multilayers, which results in the formation of multi-layered structures with a broad range of possible stochiometries. Thus, the retention of counterions within the multilayer structure reduces the impact of the entropy on the assembly of extrinsically compensated LbL films. Figure 7 shows a schematic picture describing the distribution of polyelectrolytes and counterions in both types of compensation mechanisms.

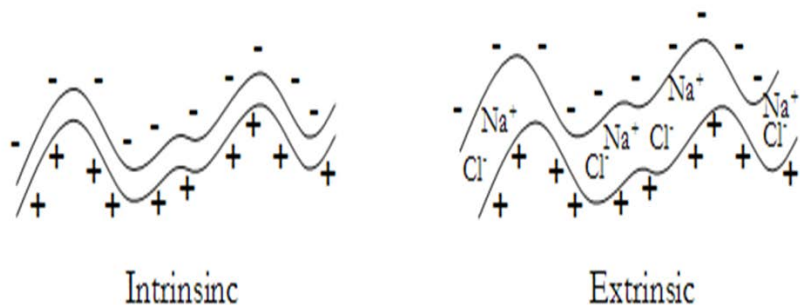

Figure 7. Scheme of the expected configuration of polyelectrolyte layers and counterions in intrinsically and extrinsically compensated polyelectrolyte multilayers. Reproduced from Ref. [107], Copyright (2009), with permission from The Royal Society of Chemistry.

The extrinsic compensation is the most frequent in polyelectrolyte multilayers [49-51, 107, 111, 118], with the intrinsic one being limited to the assembly of highly charged polyelectrolytes [107, 228, 234]. However, it is not possible to define any general rule allowing a prediction of the type of compensation expected for a specific multilayer. This is because the compensation mechanism depends on both the pair of assembled polyelectrolytes and the assembly conditions [49-51, 89, 107, 111, 118]. The ability of tuning the compensation mechanism, by changing the conditions in which the assembly is done, plays a very important role in the control of the final structure and physico-chemical properties of the LbL films [107, 111, 187]. The ionic strength of the solution is probably the most important parameter impacting on the compensation mechanism of LbL films. This is because it modify the ionic equilibrium, and as a matter of fact the effective charge density of the polyelectrolyte due to the ionic condensation phenoma [228]. The effect of the ionic strength on the charge compensation mechanism of (PDADMAC - PSS)$_n$ multilayers was demonstrated by Schlenoff and Dubas [111]. They were able to switch the multilayers from a mainly intrinsic-like compensation mechanism to an extrinsic one. This transition may be explained in terms of the role of the entropy in the assembly [107, 187]. Thus, the assembly of multilayers from solutions with low ionic strength occurs upon the release of most of the counterions to the solution, and thus an important entropy gain should be expected. On the other side, the increase of the ionic strength leads to a situation in which the release of counterions is less important, and the entropy gain does not play a significant role on the assembly. This is rationalized considering that the overall concentration of ions in the bulk is relatively high, and the release of additional counterions does not affect significantly to the energetic landscape of the system (the impact of the counterions release on the entropy of the system is rather low). Therefore, most of the counterions remain trapped within the multilayer, with such multilayers having a lacked ionic pairing between the charged groups of polyelectrolyte chains forming adjacent layers. Figure 8 shows the impact of the ionic strength on the compensation mechanism of (PDADMAC-PSS)$_n$ multilayers in terms of the dependence of monomer density, $\rho_{monomer}$, for PDADMAC and PSS layers obtained ellipsometry on the ionic strength ($I$). The higher values of the density of PDADMAC monomers in comparison to that corresponding to PSS monomer is a clear signature of the extrinsic compensation. Furthermore, this increase of the density difference with the ionic strength give an indication of the transition from an almost intrinsic compensation to a clearly extrinsic compensation for the highest values of the ionic strength.

The different values obtained for the monomer densities in polycation and polyanion layers may be explained considering that the charge compensation present an asymmetric character, i.e. the degree of extrinsic compensation in (PDADMAC - PSS)$_n$ multilayer is defined by the nature of the capping layer [107, 235]. Therefore, the higher monomer density of PDADMAC monomers leads to a situation in which the charge excess is expected to be significantly higher in PDADMAC-capped films than in PSS-capped one, and consequently an extrinsic compensation should be expected for PDADMAC layers, whereas a mostly intrinsic compensation appears for PSS layers. These differences on the compensation mechanism as function of the specific chemistry of polycation and polyanion results in important differences on the structure of the adsorbed layers. Thus, the differences on the distribution profile of counterions within the multilayers modify the osmotic stress and affect the layer structure, with the formation of PDADMAC layers with a higher hydration and swelling than PSS one. This results in higher roughness for PDADMAC-capped films than for PSS-terminated one [107, 236] (see inset Figure 8). The above discussion allows inferring that the internal charge balance in polyelectrolyte multilayers results from the combination of extrinsic sites (polyelectrolyte/counterion pairing) and intrinsic one (pairing between oppositely charged polyelectrolytes) as is schematized in Figure 9.

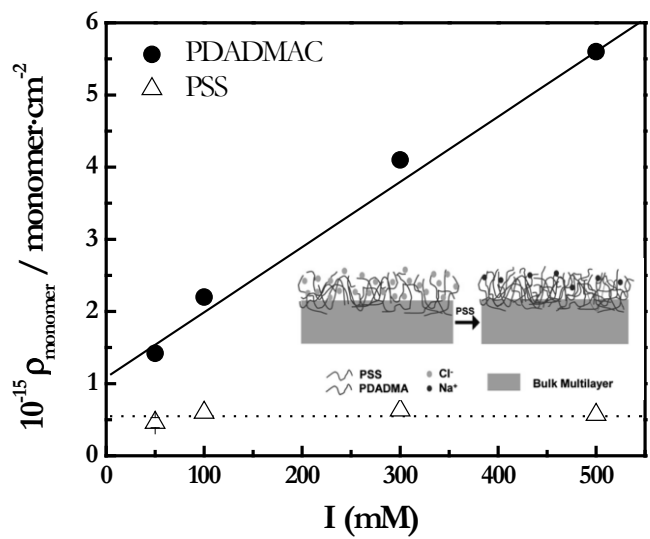

Figure 8. Surface density of monomer, $\rho_{monomer}$, for polyanion and polycation layers in (PDADMAC - PSS)$_n$ multilayers as function of the ionic strengths. The differences between the PDADMAC and PSS densities are a signature of an extrinsic-like compensation. Adapted from Ref. [107], Copyright (2009), with permission from The Royal Society of Chemistry. The inset represents the assymetrical compensation in polyelectrolyte multilayers as function of the nature of the last deposited layer. Reprinted with permission from Ref. [235]. Copyright (2012) American Chemical Society.

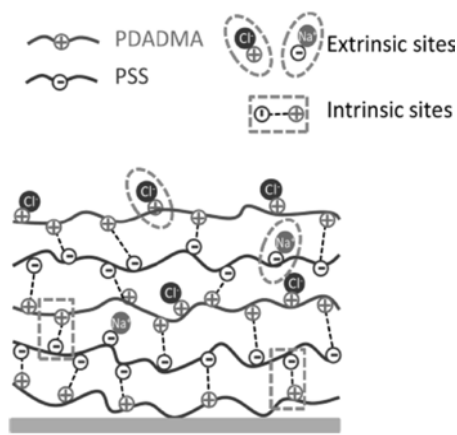

Figure 9. Scheme showing the general internal charge balance in polyelectrolyte multilayers. Reprinted from Ref. [228]. Copyright (2013) American Chemical Society

It should be expected a very different role of the enthalpy and entropy on the film assembly depending on the specific nature of the multilayers [219]. A strongly exothermic complexation is expected for multilayers presenting linear growth as (PDADMAC-PSS)$_n$ films obtained at low ionic strengths, whereas the complexation appears as endothermic for non-linear growth multilayers as (PDADMAC-PSS)$_n$ obtained at high ionic strengths [107]. Thus, assuming the common definition for the change of Gibbs energy of the system $\Delta G = \Delta H - T\Delta S$, it should be expected a a favourable role of both enthalpy $\Delta H$ (negative as result of the ionic pairing) and entropy $\Delta S$ (positive as result of the release of counterions) in linear growth multilayers, whereas for non-linear growth multilayers, even the entropic contribution continues being positive, the impact of the enthalpy is detrimental for the assembly due to its positive values. Thus, the role of the entropy and enthalpy must be considered counteractive for the assembly process [218, 237].

It should be expected that any variable affecting the ionic equilibrium, and not only the ionic strength, of the polyelectrolytes may modify the compensation mechanism (polyelectrolyte charge density, pH and solvent quality) [49, 51, 231, 238, 239]. It is worth mentioning that the ionic equilibrium is not the only contribution affecting to the overall entropy of the assembly process, with the entropy associated with the release and reorientation of hydration water [240-242], and the entropy penalty associated with the reduction of the degrees of freedom of the molecules as result to their attachment to the surface playing also a certain role [114, 243]. The latter contribution presents commonly an unfavourable effect on the assembly process of LbL. However, its impact, in most of the systems, is smaller than that of the other contributions, allowing one to neglect its role [244].

The importance role of the entropy in the assembly of polyelectrolyte multilayers could be summarised in three main reasons: (i) charge inversion has associated a high enthalpic penalty, thus an adsorption driven only by enthalpy cannot continue proceed beyond the zero net surface charge point, i.e. the compensation conditions. Therefore, the contribution of non-electrostatic interactions to the LbL assembly of polyelectrolytes ensures that the process can proceed beyond the charge compensation threshold; (ii) electrostatic interactions do not differ between intrinsic (compensation by ionic pairing) and extrinsic compensations (compensation by counterions condensation); and (iii) the possibility of building multilayers with materials presenting a reduced charge density or under conditions where the electrostatic interaction is more or less screened, e.g. high ionic strength [51, 107, 229, 245-249]. It is worth mentioning that the impact of the electrostatic and non-electrostatic contributions on the assembly of polyelectrolyte multilayers remains controvert and deserves future studies [6, 249]. The understanding of the role of the different contributions to the assembly presents a particular important because there are some cases in which a high charge density of the building blocks does not ensure the formation of multilayers, e.g. the assembly of poly(N-ethyl-4-vinylpyridinium bromide) and PMAA at pH 8.4 results in the formation of inter-polyelectrolyte complexes, even though the used polymer present a high charge density. The same occurs for the pair or that of PMAA and lysozyme at the same pH where both the building blocks have also a high charge density [250].

### 3.3. Adsorption kinetics of polyelectrolyte layers in LbL multilayers

The adsorption kinetics of polyelectrolyte layers in polyelectrolyte multilayers is probably accounted among the less explored aspects. However, several studies have shown that the understanding of the properties and structure of LbL materials requires a critical examination of the time-scales involved in the adsorption processes of the layers [107, 184, 191, 229]. This is particularly important because the stratification appearing in many polyelectrolyte multilayers may be considered the result of an arrested adsorption, in which the structure is reminiscent of a freezing at an intermediate state far from the steady state conditions [105, 106, 251]. This may be rationalized considering the long time requires for the reorganization of the polyelectrolyte chains within the multilayer structure [231]. Thus, the better stratification found for multilayers assembled using methodologies involving low contact time between the solution and the surface/multilayer, such spin-coating or spraying, than for those obtained using common dipping may be explained considering an arrest of the reorganization of the polyelectrolyte chains [124, 138].

The complex interplay of interactions involving polyelectrolyte, substrate and solvent determine the adsorption and growth of LbL films, and, in particular, the adsorption kinetics [113]. Polyelectrolytes adsorption is an almost a quasi-irreversible process, i.e. once a polyelectrolyte chain is attached to the surface it remains trapped, which can be rationalized considering that the adsorption process of a polymeric chain to a surface results from a multi-segment binding, i.e. more than one monomers attach to the surface. Thus, the desorption of an adsorbed chain requires that all the segment can be unbound simultaneously, which is difficult because during the desorption process, additional polymeric segment can attach to the surface, making almost impossible the desorption of the entire chain from the surface, at least within experimental accessible time-scales [252, 253]. The above picture provides an explanation to the impossibility of washing out a polyelectrolyte layer from the substrate upon their exposure to a solution containing a polyelectrolyte bearing the opposite charge, i.e. a counter-polyelectrolyte [107].

Most of the studies dealing with the adsorption kinetics of layers on LbL multilayers have evidenced the existence of at least two well-differentiated steps [51, 107, 184, 191, 254]. This type of kinetics can be described according to the Raposo-Avrami's adsorption model which provide a description of the time dependence of the surface concentration $\Gamma(t)$ as follows [255-257],

$$\Gamma(t) = A_1(1-e^{-t/\tau_1}) + A_2(1-e^{-t/\tau_2})^n, \tag{1}$$

with $A_1$ and $A_2$ being the amplitudes for the fast and slow adsorption steps, respectively and $\tau_1$ and $\tau_2$ the corresponding characteristic times. It is worth noting that the second term accounts for any reorganization process of the polymer chain after its adsorption, including both the diffusion within the multilayer structure and the motion of the polymer molecule in the surface plane, generally to ensure the compensation of all surface charges, with the latter being probably associated with displacements of the polymer chains on localized and uncorrelated short scales (inchworm-like motion) [109, 258]. The reorganization of the polymers chains is modeled in terms of an expression reminiscent of the Avrami's model, which is commonly applied to the description of the kinetic of polymer crystallization [259-261]. The exponent of the Avrami's term $n$ assumes values close to 1 for almost all the polyelectrolyte multilayers which allows one to rewrite Equation (1) in terms of the maximum value of the surface concentration, $\Gamma_\infty$,

$$\Gamma_\infty = A_1 + A_2 \tag{2}$$

$$\Gamma = \Gamma_\infty - A_1 e^{-t/\tau_1} - A_2 e^{-t/\tau_2} \tag{3}$$

The adsorption of a single polyelectrolyte layer during the fabrication of a LbL films involved two kinetic appearing in well-separated time scales: (i) a first fast nucleation of polyelectrolyte domains (generally below 5 minutes), and (ii) slow reorganization of the polymer chains within the multilayer (from several minutes to hours) [262]. The first step can be considered a diffusion-controlled process coupled to the adsorption of the polymer through an electrostatic or steric barrier [253, 263], whereas the second step involves all the reorganization steps driving the adsorption to the stationary state, including both in plane reorganizations of the polymeric chains and the interdiffusion along the whole multilayer structure [116, 254]. Figure 10 shows for the sake of example the adsorption kinetics of PDADMAC layer on a (PDADMAC - PSS)$_n$ film, and the discrimination of the two processes involved in the assembly process.

The results shown in Figure 10 evidence the validity of the above discussed model for providing an appropriate description of the adsorption kinetics. It is worth mentioning that the first adsorption step generally takes the system up to the 60 - 80 % of the value of $\Gamma$ corresponding to the steady state [49, 51, 89, 107]. The adsorption of most of the polyelectrolyte multilayers presents a characteristic time for the first process $\tau_1$ almost independent on the number of layers [51, 89]. However, the values of $\tau_1$ are affected by the variables that modify the assembly process (pH, ionic strength, T, etc.) in agreement with the computer simulations by Cohen-Stuart [253]. On the other side, $\tau_2$ can present complex dependences on the number of layers, making it difficult to predict the dependence for a specific multilayer [112, 116, 191]. $\tau_2$ is related to in plane reorganizations of the polymer chains, i.e. processes occurring at the multilayer surface, for systems such as (PAH - PSS)$_n$ or (PDMAEMA - PSS)$_n$ multilayers, with the characteristic time remaining almost constant with the number of layers [49, 116, 191]. However, other systems, including (PDADMAC - PSS)$_n$ or (CHI - PAA)$_n$ multilayers, present increasing $\tau_2$ values with the multilayer thickness. This suggests that the reorganization process can account for both the interdiffussion of the polymer within the multilayer tri-dimensional structure and the in plane reorganization process [51, 116, 191]. The above picture agrees with the three stage model proposed for Lane et al. [254] for the adsorption of polyelectrolyte layers on multilayers formed by PSS and poly[1-[4[(3-carboxy-4-hydroxyphenylazo)benzenesulfonamido]-1,2-ethanediyl sodium salt]. This model includes: (i) transport of the polymer from the solution to the surface, (ii) in plane reorganization of the adsorbed chains, and (iii) interdiffusion of polymer chains within the multilayer structure. Figure 11 presents a scheme including the different processes which can be involved in the adsorption of polyelectrolyte on a multilayer.

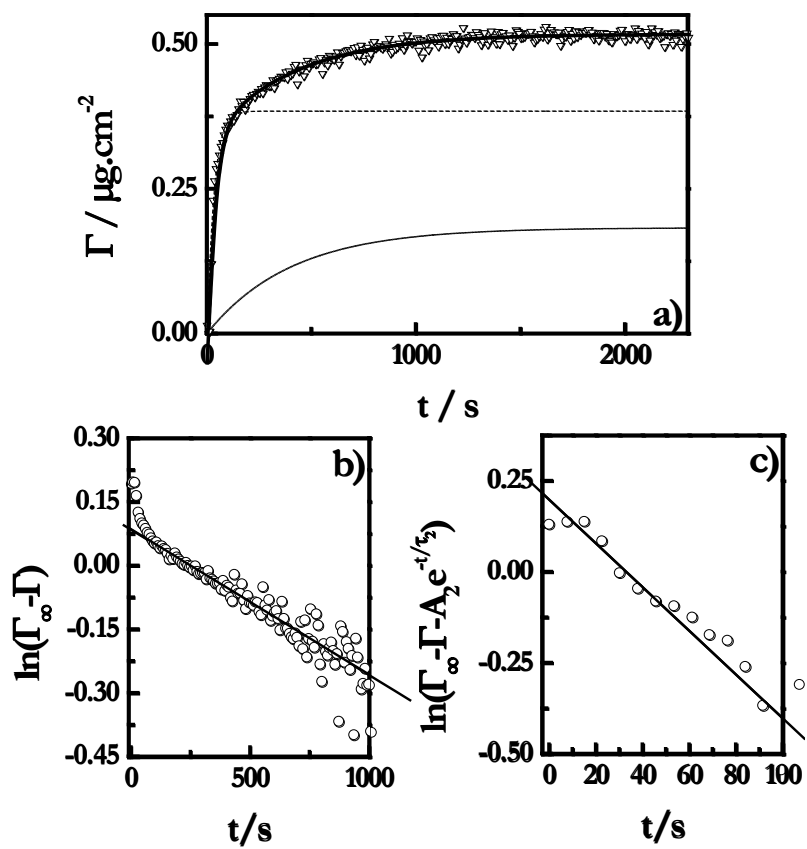

Figure 10. Adsorption kinetics, and example of their analysis, for the adsorption of PDADMAC layer on a (PDADMAC - PSS)$_n$ film. (a) Adsorption kinetics. The solid line shows the best fit to Equation (3), with the two exponential steps being evidenced: (---) first fast step and (····) second slow step. (b) Long time behavior of the adsorption kinetics plotted as $\ln(\Gamma_\infty-\Gamma)$ vs. time, with the solid line representing the fit to a straight line. (c) Short time behavior of the adsorption kinetics plotted as $\ln(\Gamma_\infty-\Gamma-A_2e^{-t/\tau_2})$ vs. time, with the fit evidenced by a solid line. Adapted from Ref. [191], Copyright (2011), with permission from Elsevier.

It is worth noting that a close look on the dependences of the adsorption times on the number of layers provides a valuable information on the determination of the possible role of the interdiffusion of the polyelectrolyte chains [116, 191], with the assembly conditions and the chemical nature of the polyelectrolyte forming the multilayer modifying significantly the adsorption kinetics [49-51, 116].

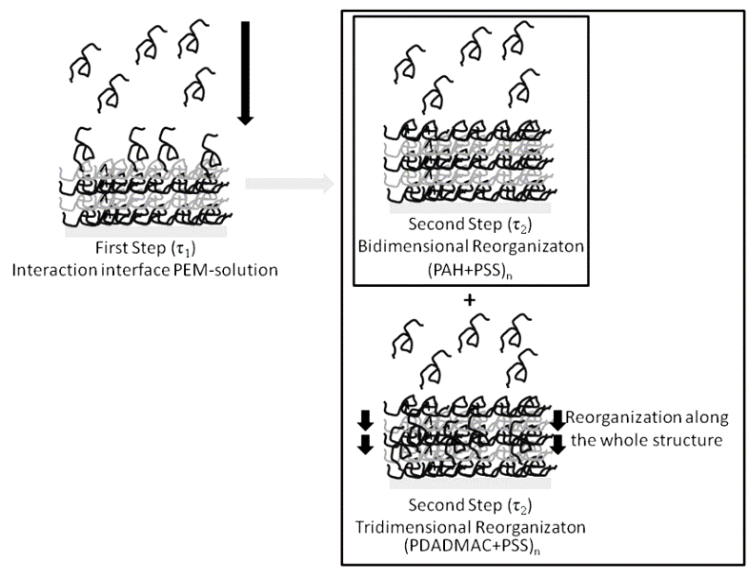

Figure 11. Scheme of the different processes involved in the adsorption of polyelectrolyte on LbL multilayers. Reprinted with permission from Ref. [116]. Copyright (2012) American Chemical Society.

### 3.4. Internal structure of polyelectrolyte multilayers

The internal structure of polyelectrolyte multilayers plays an essential role on the physico-chemical properties of the assembled systems, which makes it necessary a careful control over the different variables affecting the assembly process because they may modify the final structure of the films.

The main question to solve, when the internal structure of polyelectrolyte multilayers is concerned, is the extension of their lamellar order, i.e. whether consecutive layers are truly independent layers to enable the definition of the LbL multilayer as stratified materials [124, 138, 191]. This has been found to be strongly dependent on the nature of the assembled polyelectrolytes, the conditions used for the assembly of the films, and the protocol of fabrication (contact time and deposition method) [11, 124, 138, 191]. The general description of polyelectrolyte multilayers is based in a division of the tridimensional structure of the film into three different zones (three zone model), with a progressive transition on the multilayer structure between adjacent zones [185, 264]. The zone I corresponds to the region closest to the substrate, and it is the first region of the multilayer which is formed during the assembly of the LbL material. This regions is characterized by the alignment of the polyelectrolyte chains along the substrate surface, with its thickness remaining constant during the whole assembly process. Furthermore, the mobility of polyelectrolytes adsorbed in this region is very limited. The zone II is formed after the zone I, and presents an increasingly thickness during the assembly process. The structure of zone II resembles to that found for inter-polyelectrolyte complexes in solution [229, 264]. The outer region of the multilayer is the zone III, which maintains a constant thickness within the entire fabrication process of the multilayer, and presents a structure reminiscent to that what is expected for a free polyelectrolyte in solution [258, 265]. The three zone model was recently revisited by Singh et al. [266]. They found in (PEI-PSS)$_n$ multilayers different structural regions: (i) an inhomogeneous region near to the substrate which is associated with the influence of the interactions between the surface and the polyelectrolyte, and (ii) a homogeneous region after the deposition of a certain number of layers. Figure 12 shows a scheme of the evolution of the different zones with the multilayer growth.

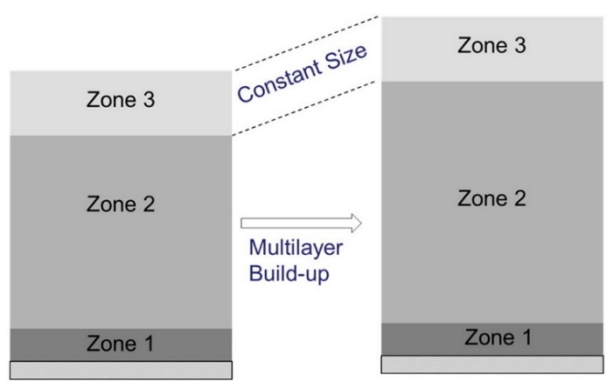

Figure 12. Scheme representing the three zone model of the multilayer structure, and its evolution with the increase of the number of layers. Adapted from Ref. [264], Copyright (2017), with permission from Elsevier.

The above picture provides a general perspective of the internal structure of polyelectrolyte multilayers. However, a more detailed structural characterization of this type of materials has been obtained using reflectivity technique, mainly neutron reflectometry (NR) and X-Ray reflectometry (XRR), and X-ray photoelectron spectroscopy (XPS) [105-107, 187, 191]. The first studies dealing with the structural characterization of polyelectrolyte multilayers paid attention to (PAH - PSS)$_n$ multilayers by combining NR and XRR [105, 106]. They performed a careful examination of the stratification of the films by NR experiments, introducing for such purpose, different sequences of layers where each certain number of layers the polyanion (PSS) was replaced by a layer of deuterated PSS (d-PSS). These experiments showed that even a certain degree of stratification can appear in the multilayers, the formation of a true lamellar structure in which independent layers may be found is far to be the real situation multilayers, existing a certain degree of interdigitation between the layers, which is related to the distance to the substrate, and generally is propagated along three or more bilayers [105, 106, 267]. Furthermore, they found that the three zone model provided an appropriate description for the correlations existing between the stratification degree of the multilayer and the thickness of the multilayer, and an increase of the roughness with the number of layers until reaching an stationary value as result of the densification occurring during the assembly process [105, 106]. The latter finding is just the opposite to that what happens for (PDADMAC-PSS)$_n$ multilayers, which present a decrease on their roughness as the growing process proceeds [268]. The difference in the roughness may be explained considering the different mechanism involved in the assembly of the polyelectrolyte layers in (PAH - PSS)$_n$ and (PDADMAC - PSS)$_n$ systems, with the interdiffusion within the film playing an important role in the latter case [184, 191], which allows one to distinguish between two different types of roughness: (i) topological roughness as that appearing in (PAH - PSS)$_n$ multilayers [105, 106], and (ii) roughness associated with the assembly process as that found in (PDADMAC - PSS)$_n$ multilayers [268]. It is worth mentioning that analogous differences has been reported by Guzmán et al. [191].

The differences between (PAH - PSS)$_n$ and (PDADMAC - PSS)$_n$ multilayers go beyond the impact of the assembling processes on the roughness of the films [105-107, 184, 187, 191] as was pointed out by Guzmán et al. [107, 187, 191]. They showed, combining NR, XRR and XPS, that the stratification of (PDADMAC-PSS)$_n$ was completely absent, independently of the assembling conditions used for the building of the multilayers, which may be explained considering the differences of the times used for the layer assembly, with longer time used for the assembly of (PDADMAC-PSS)$_n$ multilayer that for (PAH - PSS)$_n$ one (adsorption of the layers to reach the steady state conditions, around 1 hour [107], *vs.* adsorption during 15 – 20 minutes [105, 106]). This suggest that the structure of polyelectrolyte multilayers may be affected by a dynamic constrain, which favors the formation of films with a partial stratification as result of an arrested adsorption as was predicted by Panchagnula et al. [251] on the basis of molecular dynamics simulations. It is worth mentioning that deepening in the impact of the time on the stratification of LbL layer, Guzmán et al. [191] found that the stratification of LbL film depends on the adsorption kinetics of the layers, with (PDADMAC - PSS)$_n$ multilayers where the interdiffusion presents a central role on the assembly no showing any signature of stratification, whereas (PAH - PSS)$_n$ multilayers evidence a time dependent stratification [191]. The temporal evolution of the structure of the multilayers has been also observed by Ge et al. [269] using Vibrational Sum Frequency Generation Spectroscopy. The impact of the time used for the layer assembly on the lamellar order of polyelectrolyte multilayers has been recently confirmed by Selin et al. [270]. They found that multilayers of poly(methacrylic acid) (PMAA) as polyanion and quaternized poly(2-(dimethylamino)ethyl methacrylate) (QPC) as polycation present an enhanced intermixing of the layers as the adsorption time increases. This was evidenced from the evolution of the neutron scattering length density profiles with the adsorption time: (i) short adsorption times result in a layer intermixing which occurs only on the most external layers, and (ii) long adsorption times result in a intermixing within the entire multilayers. Therefore, the adsorption time can be identified as the driving force leading to the increase of the thickness of the region within the intermixing occurs. A recent study has shown that the thermal annealing can be used for the destruction of the lamellar structure [271]. This is the result of two simultaneous process which impact on different regions of the film: (i) densification, and (ii) degradation. The impact of the annealing with salt on the structure of (PDADMAC - PSS)$_n$ multilayers was found to be similar to the thermal annealing [272]. The absence of a true stratification in polyelectrolyte LbL materials makes it is necessary to redefine this type of systems as solid supported inter-polyelectrolyte complexes, existing two types: (i) partially stratified, and (ii) non-stratified[273].

Many technological fields based on LbL materials require stratified systems. This can be solved using the approach proposed by Gilbert et al. [274]. They design a methodology enabling the fabrication of stratified polyelectrolyte multilayers, including, in the multi-layered structure of the film, different single layers which blocks the interdiffusion of the other polymers. It is worth mentioning that the absence of such layers favours the formation of homogeneously interdigitated films.

## 4. Controlling the assembly of Layer-by-Layer multilayers: parameter modifying the assembly

The above discussion evidenced that the LbL method is not a true methodology for coating fabrication, with this methodology being a self-limiting adsorption process which drives the formation inter-polyelectrolyte complexes supported, in most of the cases, by a template. The control of the properties and structure of such materials requires a careful examination of the bath conditions (pH, added salt, concentration), and the properties of the building blocks and templates [113, 275].

### 4.1. Physico-chemical properties of the template

The discussion of Section 2 evidenced that that polyelectrolyte multilayers may be assembled in a broad range of substrates with different chemical nature, morphology and size [114]. The specific chemical nature of the surfaces, and, in particular, their surface charge density are critical parameters for the assembly of LbL materials onto its surface [11]. This is because the suitability of a specific substrate as template is correlated to the specific nature of the interactions occurring between the polyelectrolyte and the substrate. This is generally related to the hydrophilicity/phobicity characteristics of the template, its chemical nature and the density of the charged groups existing in its surface, as well as to its roughness, porosity and the presence of impurities [276]. This is because such parameters present a critical impact on the homogeneity and the stability of the obtained multi-layer films, or at least on the corresponding to the layers which interact stronger with the substrate, i.e. the first deposited layers [185, 277-281]. Furthermore, the interactions between the substrate and the initially deposited layers can control the specific growth mechanism as was discussed above [182, 218, 219].

### 4.2. Concentration of the solutions of polyelectrolytes

The fabrication of LbL materials occurs after the exposure of the substrate to be coated to a solution containing the material to be assembled, which requires the use of solutions with a concentration high enough to enable the deposition of stable layers and to ensure the charge inversion in those systems where the electrostatic contribution is critical [184, 191, 210]. Therefore, the minimum concentration required for the assembly of LbL films will depend on the solubility and charge density of the assembled blocks, and it would be expected that the increase of the concentration of the solutions beyond a certain threshold value present a rather limited impact on the assembly process [191]. Figure 13 shows the polycation concentration dependence of the thickness per bilayer of the linear growth (PDMAEMA-PSS)$_n$ multilayers in which polycations with different charge densities are included. It is clear the increase of the thickness per bilayer with the polycation concentration up to reach a threshold concentration value.

The increase of the polyelectrolyte concentration leads to the formation of thicker multilayers [49, 191]. This is rationalized considering the formation of fuzzier layers, with a higher proportion of loops and tails protruding to the solution, as result of the competence of the polymers chains for the binding points of the surface. Furthermore, the polymer concentration may affect significantly to the growth mechanism and the intermolecular association within the multilayers as was evidenced by Shen et al. [282]. They showed that for (PLL-HA)$_n$ multilayers, the increase of the HA concentration above a threshold value (around 2 mg/mL) results in a transition from a linear growth to a non-linear one. Guzman et al. [49] and a Garg et al. [283] show similar dependence for (PDMAEMA-PSS)$_n$ multilayers and for those formed by PAH and Poly[1-(*p*-(3'-carboxy-4'-hydroxyphenylazo)benzenesulfonamide)-1,2-ethandiyl], respectively.

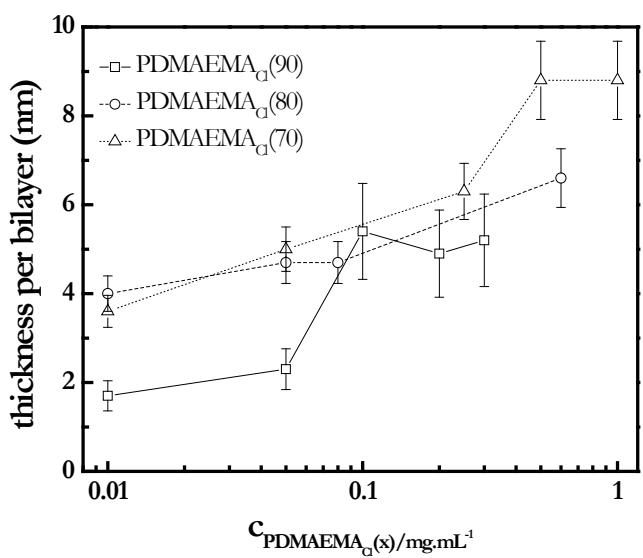

Figure 13. Polycation concentration dependences of the average thickness per bilayer of (PDMAEMA(*x*)-PSS)$_n$, where *x* indicates the weight fraction of the charged block in the copolymer. Adapted with permission from Ref. [49]. Copyright (2010) American Chemical Society.

### 4.3. Nature of the building blocks: chemistry, charge density and molecular weight

The chemical nature of the polyelectrolytes and, in particular, their charge density, together with the assembly conditions play a central role in the control of the assembly of the LbL multilayer. This is explained considering their impact on the complex balance of interactions existing during the assembly process, and the ability of the assembly blocks for being assembled trough electrostatic interactions or any other specific interactions. A careful examination of the role of the interactions is essential for tuning the thickness, structure and properties of the LbL multilayers [113, 212]. This is especially important, when the deposition of the initially deposited layers is concerned because the attachment of such first layers to the substrates is critical on the stability of the obtained multi-layered structure [113].

The first aspect to consider in relation to the impact of the chemical nature of the building blocks is the role of the hydrophilic/hydrophobic balance of the polymers. This is because it determines both the interactions between the polyelectrolyte and the swelling degree of the layers [284]. It has been demonstrated that the thickness of the layers increases with the hydrophobicity of the polyelectrolytes, e.g. (PDADMAC + PAMS)$_n$ (with PAMS being poly(acrylamidesulfonic acid)) < (PAH + PSS)$_n$ < (PDADMAC + PSS)$_n$ [285]. This is explained considering the penalty associated with the solvation of the chains [118]. Furthermore, the poorer solubility of hydrophobic blocks in polar solvents, as water, is associated with the depletion of the material from the solution, which contributes to its deposition on the multilayer [50, 240, 243, 263, 286]. The flexibility of the polymer chains also impact decisively on the assembly of LbL films because it modifies the conformation of the layers, and as matter of fact the interaction between the chains in solution and the layers [113, 116, 191, 212].

The charge density of the building blocks is also an important parameter on the fabrication of polyelectrolyte LbL multilayers, mainly because the interactions between charges plays a very important role on the fabrication of this type of materials. Therefore, it is expected the existence of a minimum charge density threshold for enabling the building process [114, 115, 246, 287, 288]. The threshold charge density is a reference state below which the formation of stable multilayers is not possible. This is explained considering that below a threshold value of the charge density, the adsorption of the polyelectrolyte is relatively weak, and thus the exposure of the adsorbed to a solution containing an oppositely charged polyelectrolyte results in a removal of the weakly bound layer, resulting in the formation of complexes into the bulk. This is favored because the enthalpy change associated with the electrostatic interaction cannot compensate the entropy gain due to the release of the weakly adsorbed polymer from the surface, yielding in the desorption of the chains [118, 175]. It is worth mentioning that the charge density threshold may be shifted by adding salt to the polyelectrolyte solutions, i.e. enhancing the role of the hydrophobic interactions [289]. The importance of the charge density of the building blocks on the assembly of polyelectrolyte multilayers was evidenced by Glinel et al. [115]. They studied multilayers between PSS and statistic copolymers of (diallyl-dimethylammonium chloride) and (N-methylformamide) with different molar ratio between the ionic and non-ionic blocks, and found three different zones for the growth of these multilayers: (i) for copolymers with the lowest charge densities (below 10%), no multilayer growth was found; (ii) for copolymers with charge densities in the 10-40% range, the thickness of the multilayers increases as the charge density of copolymers decreases, and (iii) for copolymers with charge densities above 40%, the thickness of the multilayers remains almost unchanged with the charge density of the copolymer.

It is true that a minimal charge density is important for the LbL assembly because the electrostatic self-assembly of polyelectrolyte is the result of two counteracting forces: (i) electrostatic attraction between the polyelectrolyte and the oppositely charged substrate, and (ii) solubility of the polyelectrolyte. The increase of the charge density makes it less favorable the adsorption of the polyelectrolyte, which leads to a situation in which the thicker multilayers are obtained for polymers presenting a charge density between the lower charge density threshold and the maximum nominal charge density of the polyelectrolyte. The dependence of the adsorbed amount on the charge density of the polyelectrolyte can be explained as result of the balance between the above mentioned forces [288]. Therefore, the variation of the thickness for charge densities above the charge density threshold value can be understood assuming that the increase of the charge density of the copolymer drives the system from a charge-dependent "Debye Hückel" regime to a charge-independent "strong-screening" regime [115]. It is worth mentioning that the analysis of the impact of the charge density on the adsorption of polyelectrolyte also requires a careful examination of its distribution within the polymeric chain, e.g. the adsorption of block copolymers, presenting at least a strongly charge block, can be significant even for polymers with a reduced charge density (around the 10% of the number of monomers) [238].

A last but not least aspect to consider on the impact of the polyelectrolyte nature on the assembly of multilayers is the role of the molecular weight of the chains in the assembly process. The high and low molecular weight concepts in polyelectrolyte multilayers should be analyzed in relation to their impact on the multilayer growth [149, 290]. In general, the assembly of low molecular weight polyelectrolytes (around $10^3$ Da) leads to the stripping off of a polyelectrolyte layer upon its exposure to a polyelectrolyte of opposite charge, which hinders the multilayer growth. On the other side, the increase of the molecular mass of the assembled polyelectrolytes enhances the deposition of the multilayers. This may be easily understood considering that the adsorption-desorption is a kinetically-controlled process and the increase of the molecular weight of the adsorbing species slows down the exchange of adsorbed polymer between the multilayer and the solution. This allows concluding that the impact of the molecular weight on the multilayer growth is the result of a complicate balance between two counteracting forces: (i) thermodynamically-driven stripping resulting in the formation of inter-polyelectrolyte complexes in solution, and (ii) kinetically-driven sticking of the polymers resulting in multi-layering [149].

The impact of the molecular weight of the assembled polyelectrolytes has been recently revisited by Towle et al. [117]. Their study was focused on the assembly of (PLL-PAA)$_n$ multilayers using polyelectrolytes with a broad range of molecular weights, and they found an increase of the adsorbed amount and rigidity of the obtained films with the molecular weight of the assembled polymer. However, they also found that the specific charge density distribution and the stripping phenomena results in a non-monotonic variation of the surface roughness and the surface energy of the films.

### 4.4. Methodological aspects: drying, rinsing and contact time

The analysis of the role of different methodological aspects, including the contact time between the solution and the multilayer, as well as the rinsing and drying steps between the adsorption of adjacent layers, on the building of polyelectrolyte multilayers is essential.

The role of the contact time on the assembly of LbL multilayers was introduced in Section 3 where the impact of this aspect on the fabrication process was clearly stated. This is especially true when multilayers involving interdiffusion are concerned. The use of long adsorption times results, in most of the cases, in a weakening of the lamellar structure of the multilayers due to the internal reorganization of the polyelectrolyte chains within the tridimensional structure of the multilayer [191]. However, the use of short contact times between the solution and the multilayer results in a kinetical arrest of the interdifussion [251], which leads to a certain degree of lamellar structure in the film [105, 106]. Thus, it is possible to assume that the stratification of the polyelectrolyte multilayers is correlated to the contact time, and hence the fabrication of stratified films makes it necessary the use of short adsorption times. This agrees with the results by Kharlampieva et al. [138] about the impact of the methodology used for the assembly of (PAH - PSS)$_n$ multilayers on the internal structure of the multilayer. They found that the use of fabrication techniques requiring a low contact time between the solution and the multilayer (e.g. spray-assisted deposition or spin-coating: 30 – 180 seconds) results in multilayers with a better stratification than those obtained using methodologies where the contact time is long enough for enabling the interdiffusion of the polymer chains (dipping method). This scenario agrees with the better stratification reported by Félix et al. [124] for other multilayers as the contact time was reduced. Furthermore, the reduction of the contact time can allow an improvement on the homogeneity of the assembled films as result of the limitation of the polymer diffusion [146]. Thus, the fabrication of LbL materials present an important kinetics control [291].

The rinsing between the deposition of adjacent layers presents a key importance because it allows removing the excess of deposited materials which is adsorbed onto the surface through weak intermolecular interactions. This is especially important when the deposition of polyelectrolytes is analyzed because in absence of rinsing steps, the exposure of a multilayer capped with a polyelectrolyte to a solution of a second polyelectrolyte bearing the opposite charge can result in the formation of inter-polyelectrolyte complexes in solution (thermodynamically favored), with the precipitation of such complexes onto the multilayer distorting the structure and properties of the films [11, 292-294]. It is worth mentioning that the introduction of rinsing steps in the assembly process affects very differently to polyelectrolytes [175]. The layers of strong polyelectrolytes remain almost unaltered after their exposure to rinsing solution as result of their almost irreversible attachment through strong electrostatic interactions. On the other side, the weakness of the binding strength of layers formed by weak polyelectrolytes may result in an easier removal upon exposure to a rinsing solution [11, 295]. It is worth noting that recent studies have suggested the important role of the time used for the rinsing step to ensure a correct removal of the excess of polyelectrolyte. This is even more important when the deposition of LbL films onto heterogeneous surfaces is analyzed [291].

Another methodological aspect to be considered on the assembly of LbL layers is the drying of the film between two consecutive deposition steps. Different studies have reported the essential role of the drying for tuning the structure and physico-chemical properties of polyelectrolyte multilayers [11, 262] This is because the LbL method is a wet methodology, and the drying may modify the hydration and swelling of the adsorbed layers which may modify the structural organization of the multilayer [11, 262]. These modifications may affect to the growth mechanism of the films, and even the drying of the films may hinder the adsorption of additional layers, stopping consequently the propagation of the assembly process [134]. Raposo et al. [257] pointed out that the drying between the adjacent layer was mandatory in (POMA - PVS)$_n$ multilayers (with POMA and PVS being poly(o-methoxyaniline) and poly(vinylsulfonic acid), respectively) for ensuring an optimal growth of the films, with the methodology followed for the drying process affecting to both the multilayer structure and the properties of the obtained film. This is clear from the different structures of (PAH - PSS)$_n$ multilayers observed for multilayers dried under ambient air and under a nitrogen stream, with the latter being more disordered [296, 297]. It is worth noting that the impact of the drying on the assembly of LbL materials is strongly correlated to the assembly protocol, which may modify the stability of the films and their applicability.

### 4.5. Solvent polarity

Most of the films obtained by the LbL method are generally obtained from polyelectrolyte solutions in water. This is because water is a good solvent for almost all the polyelectrolyte, which is favourable for a deposition approach based on the contact of solutions containing the material to be assembled and the substrate. Furthermore, the use of water allows minimizing the toxicity associated with the use of most of the organic solvent. However, in some case can be interesting the use of organic solvents with different polarities on the assembly of LbL materials to tune the interactions and conformations of the building blocks, and as matter of fact the structure and properties of the obtained films. Among the organic solvents for controlling the assembly of LbL films can be accounted ethanol, dimethylformamide, dimethylsulfoxide or chloroform [118, 298-300]. The use of such solvents provides the bases for decreasing the role of the electrostatic interactions on the deposition, tuning the control of the assembly toward the dispersion forces and hydrogen bonds [11].

It was shown that the assembly of (PDADMAC - PSS)$_n$ multilayers from aqueous solutions including increasing ethanol weight fractions on the deposition of thicker films [239]. This may be rationalized that the decrease of the dielectric constant of the solvent associated with the addition of ethanol worsens the solubility of the polyelectrolytes because it screens the electrostatic interactions, favouring their depletion from the solution which results in an enhanced deposition. Similar results have been observed by the use of chloroform or dimethylformamide to modify the interactions involved in the assembly of (PAH - PSS)$_n$ multilayers [299, 300]. Therefore, the worsening of the solvent quality for a specific building block may be favourable for enhancing the deposition [301-303].

### 4.6. Ionic strength: concentration and nature of the supporting electrolyte

The change of the ionic strength, $I$, of the solutions can strongly affect the complexation process between the polycations and the polyanions during the assembly of LbL films [110]. This is rationalized considering the incorporation of additional counterions to the system modifies the balance between the enthalpic and entropic contributions to the assembly process as was discussed in Section 3. Furthermore, the additional counterions screen the charges along the chains, decreasing their effective charge, and consequently the inter- and intra-chains repulsion between the polyelectrolytes that results on a conformational transition from a rod-like conformation to a more coiled one. This results in a worsening of the solubility of the polyelectrolyte and, in most of the cases, in an enhanced deposition and a change of the multilayer structure [107, 110, 111]. This is clear from the results in Figure 14 where an increase of the film thickness with the ionic strength is observed for (PDADMAC - PSS)$_n$ and (PAH - PSS)$_n$ multilayers [116].

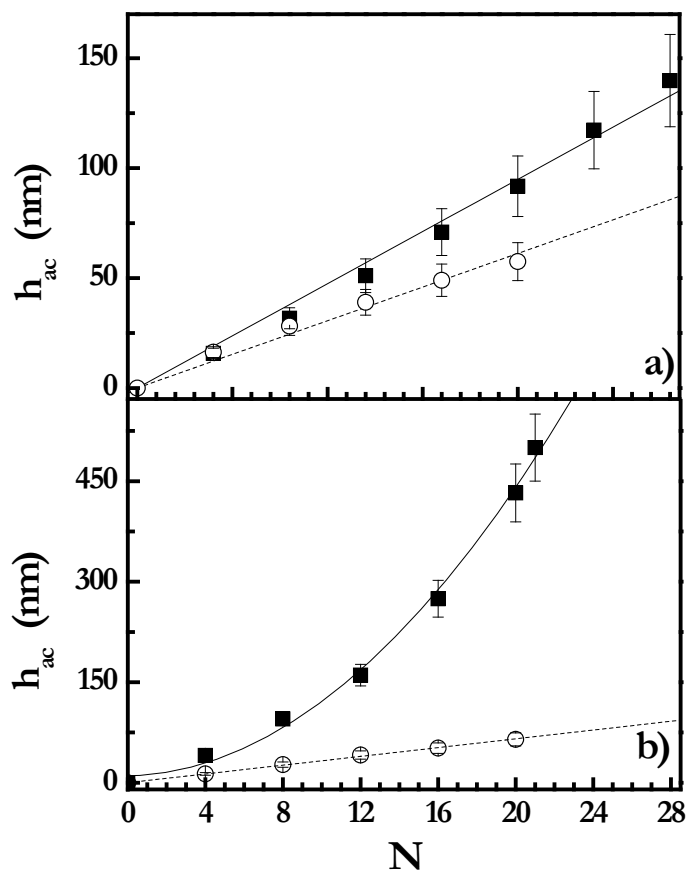

Figure 14. Effect of $I$ on the growth of (PDADMAC-PSS)$_n$ (■) and (PAH-PSS)$_n$ (○) multilayers as $h_{ac}$ (thickness obtained using a Dissipative Quartz Crystal Microbalance) vs. number of layers at $I = 100$ mM (a) and $I = 500$ mM (b). The lines are guides for the eyes. Adapted with permission from Ref. [116]. Copyright (2012) American Chemical Society.

It is worth mentioning that there is an upper ionic strength threshold value above which the thickness of the multilayer is not modified, and even for specific polyelectrolyte pairs the assembly process can be completely hindered. This is explained considering a high level of screening of the polyelectrolyte charges, which limits the adsorption of the chains via electrostatic interactions [304]. Furthermore, there are systems where a high increase of the ionic strength can result on the deconstruction of the multilayers. This is critically correlated to degree of ionic pairing occurring in the multilayers [305], which is rationalized assuming a weakening of the complexation process as result of the reduced charge density of the polyelectrolyte chains, resulting in less stable multilayers. The threshold value of salt concentration resulting in the deconstruction of the multilayers (stable-unstable) depends strongly on both the type of salt and the charge density of the polyelectrolytes [219].

The above discussion evidenced that together with the salt concentration, the type of supporting electrolyte can affect decisively on the properties and growth of polyelectrolyte multilayers. This may be described in terms of the ion specific effects accounted by the Hofmeister series, which ranks the ions in terms of its ability to order or disorder the water around the molecules [306, 307]. This modifies the swelling degree and hydration of the multilayers, which results in different modification of the properties and structure of the multilayers [308-310]. It is general that small ions (the so-called cosmotropic ions, water structure makers), such as $F^-$ or $Li^+$, present a weak binding to the multilayer as result of their small polarizability. Thus, they are easily excluded from the multilayer, resulting in the formation of thin films with a reduced hydration and roughness. On the other side, big ions (the so-called chaotropic ions, water structure breakers), such as $I^-$ or $Cs^+$, present a stronger binding to the multilayer due to its larger polarizability. This results in a stronger screening of the polyelectrolyte charges, i.e. a more extrinsic compensation, and consequently the polymers adopt a more coiled conformation, which result in an increase of the thickness and roughness of the obtained films [308, 311, 312]. Therefore, it is possible to assume that the hydrophobicity degree of the ions modifies the thickness of the multilayers, i.e. thickness increases with hydrophobicity of the ions, with cations impacting less on the growth of polyelectrolyte multilayers than anions [307, 313]. It is worth noting that the contribution of the ionic specific effects appears for ionic strength values exceeding a threshold value (around 100 mM), with the common electrostatic interactions govern the assembly of multilayers below such threshold [311]. The analysis of the building of (PDADMAC - PSS)$_n$ multilayers in presence of different sodium salt evidenced dependences for the thickness and roughness of the multilayers according to the following series: $Br^- > NO_3^- > ClO_3^- > Cl^- > BrO_3^- > HCOO^- > F^-$ [308, 309]. Figure 15 shows the dependences of the thickness on the number of layers for (PDADMAC - PSS)$_n$ multilayers deposited from solutions containing sodium salts presenting different anions, and a scheme of association of two different counterions with the deposited polyelectrolyte layers.

The effects associated with the differences on the hydrophobicity of the cations are less important than those corresponding to the anions. However, Long et al. [313] found that multilayers the thickness of the of (PDADMAC - PAMS)$_n$ multilayers can be tuned using ionic salt containing different cations during the assembly. The results show that the thickness of the multilayers showed the following dependence on the nature of the cation: $Li^+<Na^+<K^+$. Similar results have been found for (PAH-PSS)$_n$ multilayers, where the high strength of the interaction between $Cs^+$ and PSS leads to the formation of unstable films [314]. Therefore, the ionic specific effects may drive the transition between linear and non-linear growth mechanisms or instabilities on the multilayer growth as the ion size increases [308, 311].

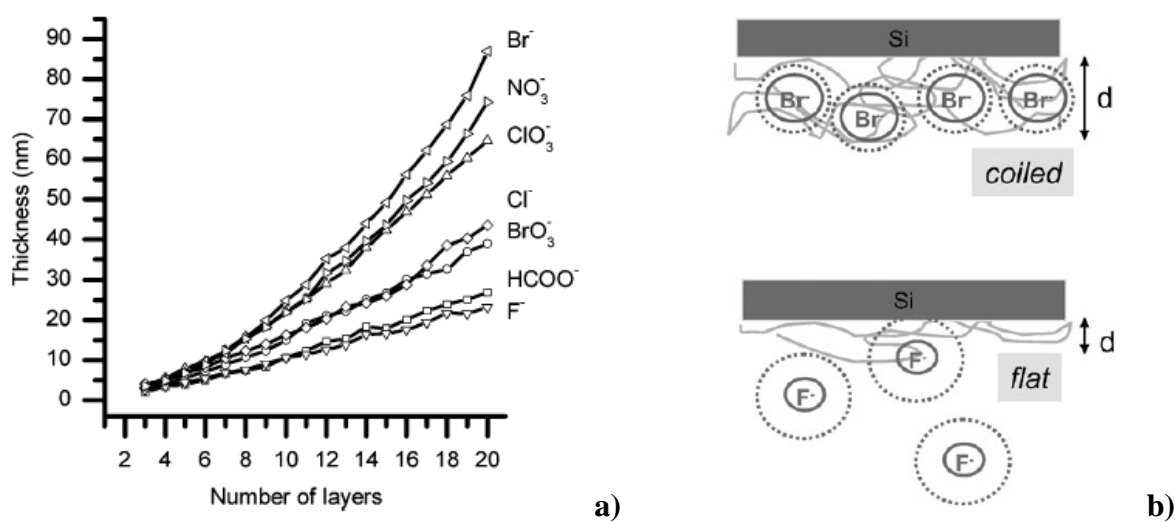

Figure 15. (a) Dependence of the film thickness as was determined by ellipsometry on the number of layers for (PDADMAC - PSS)$_n$ multilayers deposited from solution containing a concentration 0.1 M of sodium salt with different anions. Adapted with permission from Ref. [308]. Copyright (2004) American Chemical Society. (b) Scheme representing the effect of two counterions with very different effect on the multilayer growth. Reproduced from Ref. [113], Copyright (2006), with permission from the PCCP Owner Societies.

It is worth mentioning that the Hofmeister series only provides an appropriate description on the effect of univalent salts on the thickness, degree of swelling, and extent of layer interpenetration of conventional polyelectrolyte multilayers [315]. The situation appears more complex when multivalent salts are incorporated. Multivalent ions can induce the formation of intra- or interchain bridge of the polyelectrolyte [242, 316] as was stated by Dressick et al. [315] for (PAH

+ PSS)n films, with the formation of bridges between amino groups of the PAH chains limiting the chains packing. This results in the formation of highly porous supramolecular films.

**4.7.  pH**

The modification of the ionic equilibrium, and consequently of the effective charge of the building blocks by changing the pH presents a big importance on the assembly of polyelectrolyte multilayers, and in particular, when the assembly of weak polyelectrolytes, i.e. polyelectrolytes which a charge density tunable for the pH of the solutions, is concerned. The pH affects the degree of ionization of the polyelectrolyte chains, with the decrease of the latter driving a reduction of the polyelectrolyte solubility and their depletion from the solution. This is expected to enhance the deposition and increase the fuzziness of the obtained layers [51]. The understanding of the ionic equilibrium in the assembly of charged blocks is essential for using the pH as parameter for controlling the multilayer building. A simplified description of this ionic equilibrium is

$$Pol^-M^+_{(m)} + Pol^+A^-_{(aq)} \longleftrightarrow Pol^-Pol^+ + M^+_{(aq)} + A^-_{(aq)}$$

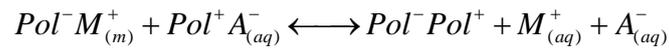

with *m* and *aq* being referred to the multilayer and aqueous phases, respectively, and *M* and *A* indicating the counterions of each polyelectrolyte and *Pol* with the positive and negative superindexes being the polycation and polyanion, respectively.

It is worth mentioning that the bases underlying the effect of the pH changes on the assembly of multilayer involving weak polyelectrolytes can be understood in a similar way to that discussed for the effect of the ionic strength [51, 119]. The effect of the pH can be considered very important in multilayers involving PAH, PAA or different biopolymers. Guzman et al. [51] evidenced the strong effect of the pH on the ionization and solubility of PAA and CHI chains, which affect to the thickness of (PAA - CHI)n multilayers. Furthermore, the modification of the pH can be used to induce a transition between different growth mechanisms as shown Bieker and Schönhoff [119] in (PAH + PAA)n multilayers. Gong [317] showed that the surface morphology of multilayers of PDADMAC and poly(4-styrenesulfonic acid-*co*-maleic acid) can be modified by the assembly pH, with the roughness of the layers of PDADMAC (strong polyelectrolyte, non-pH sensitive) remaining unchanged with pH, whereas that of the polyanion layers, having a 50% of pH sensitive monomers (maleic acid monomers), increases with the pH as result of the worsening of the solubility of the copolymer which leads to a fuzzier structure of the copolymer-capped multilayers.

### 4.8. Temperature

The importance of the temperature on the assembly of polyelectrolyte multilayers has received less attention than other parameters. However, its impact on the control of the solubility of the polyelectrolyte can be even stronger than that reported for the pH or the ionic strength [120]. This is clearly analyzing the particular case of (PAH - PSS)$_n$. multilayers where the temperature is the most effective way to induce their non-linear growth. The transition between linear and non-linear growth mechanism can be also induced on (PDADMAC - PSS)$_n$ multilayers by temperature changes [31, 185, 189], with the rate of exponential build up following a temperature dependant law *h(N+1)=h(N)e$^{\beta N}$*, where *h* indicates the thickness and *N* the number of deposited bilayers. *β* is the growth exponent which follows an Arrhenius-like dependence on 1/*T*. This gives an indication of the enhanced ability of the polymer for overcoming activation barrier with the increase of *T* [120]. Thus, the increase of the temperature weaken the ionic pairing in the multilayers, making easier the mobility of the chains within the multilayer [219, 318].

### 5. Physico-chemical properties of LbL multilayers

The discussion of the previous section has been devoted on the main physico-chemical aspects associated with the assembly of LbL multilayer. However, the understanding of the behavior of LbL materials is a challenge for both fundamental and applied science. Furthermore, the potential use of polyelectrolyte multilayers for a specific application requires a careful examination of their physico-chemical properties, and the procedures allowing one to tune such properties at will for obtaining materials with tailored properties. This section includes a brief discussion of some of the most important properties for the technological application of LbL films.

### 5.1. Water content: hydration and swelling

The water included within the multilayers is essential in the dynamic behavior of the films and their structure [49-51, 89, 107, 108, 311, 319-326]. This is the result of the impact of the water on the local molecular interactions, especially those related to the complexation between the polyelectrolytes during the assembly process [321, 325, 326]. It is worth mentioning that the water trapped in LbL materials affects to two different aspects of the multilayers: (i) hydration and (ii) swelling [320, 325, 326]. The dependence with the number of deposited layer of these two types of water was found to be the opposite [325, 326].

Lösche et al. [106] using neutron reflectometry quantified a water content in (PAH-PSS)$_n$ multilayers close to 40% of the total weight of the multilayer which corresponds to 8 water molecules per polyanion-polycation pair. For the case of (PDADMAC-PSS)$_n$ multilayers, the number of water molecules per PDADMAC-PSS pair was found to be slightly lower (about 6), which is ascribed to the higher hydrophobicity of PDADMAC than PSS [110]. Most of polyelectrolyte multilayers present a relatively high water content, commonly between the 20 and 80 % of the total weight of the films [49, 51, 107, 325, 326]. This is tunable by changing the conditions used during the assembly, mainly the charge density of the assembled blocks [321]. The high weight fraction ($X_w$) of water on polyelectrolyte multilayers is essential for the properties of these systems, impacting decisively on aspects such as the encapsulation and release of active compounds from the assembled LbL materials. The water content of the films is strongly dependent on the multilayer thickness, and decreases as the thickness or the number of deposited layers increase as result of the material densification (see Figure 16) [107, 327]. It is worth mentioning that hydration water is asymmetrically distributed within the multilayer, with an even-odd dependence of the average water content of the multilayer on the nature of the outermost layer [312, 328]. This asymmetrical distribution of the hydration water and the key role of the last deposited layer in the hydration properties were confirmed by Ghostine et al. [236] and Lehaf et al. [235]. They show that the amount of water in (PDADMAC - PSS)$_n$ multilayers is higher for PDADMAC-capped films than for those terminated in PSS. This is explained considering the different ability of the polyelectrolytes to associate water which agrees with the oscillation in the water mobility within the multilayer obtained by Schwarz and Schönhoff [322] using Nuclear Magnetic Resonance (NMR).

The high water content trapped within the multilayer during the process of assembly makes that this films can be considered as an example of gel-like materials, i.e. hydrated polyelectrolyte multilayers are plasticized materials [51]. This may be of interest for different applications in which the diffusion of small molecules along the film is required, e.g. drug delivery systems [113, 325, 326].

The role of water in multilayers is not only the hydration and plastification of the films, playing an essential role in the degree of swelling of the materials [51, 107, 329]. This is strongly affected for the environmental conditions (e.g. relative humidity) and ionic strength of the solutions [330]. This is again a result of the effect on the ionic strength on the ionic pairing between adjacent layers, and as matter of fact on the ionic cross-linking along the multilayer, which is expected to be critical on the control of the mechanical properties of the LbL materials [50, 51]. It is worth mentioning that the swelling of the multilayers is strongly affected for the interactions involved within the films, thus the nature of the building blocks will affect to the swelling degree of the multilayers [51, 240, 331].

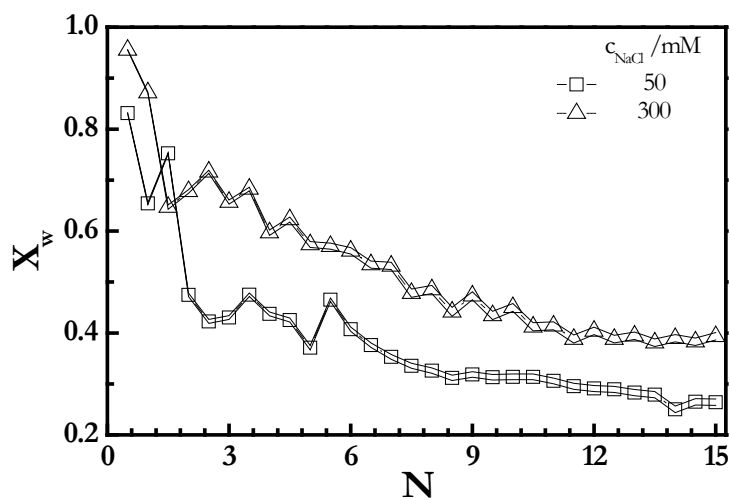

Figure 16. Water content dependence on the number of bilayers for (PDADMAC - PSS)$_n$ multilayers deposited from solutions with different ionic strength. Lines are guides for the eyes. Adapted from Ref. [107], Copyright 2011, with permission from The Royal Society of Chemistry.

### 5.2. Mechanical properties

The mechanical properties of the LbL films play a central role for designing functional materials due to their importance on the control of the stability and the response of the materials against external perturbations [332, 333]. The mechanical properties are linked to both the hydration degree of the multilayers, and the specific interactions occurring within the multilayers. It was stated above that the high water content of the multilayers leads to their softening, and the mechanical response of the LbL films appears to be reminiscent to that found for gel-like or rubber-like materials, with viscoelastic modulus in the MPa range [51, 107]. However, the impact of the water content on the mechanical response of the multilayers makes possible to find dry films with a rigidity reminiscent of the behaviour of glass-like material [51, 107, 334, 335]. Pavoor et al. [336] show for (PAH – PAA)$_n$ multilayers that multilayers obtained under conditions ensuring a high charge density of the polyelectrolyte chains present a high ionic pairing, excluding water from the multilayers. This result in high values of the Young modulus obtained using quasi-static nano-indentation (up to 10 GPa for dry films, and two order of magnitude lower for wet films). On the other side, the deposition of chains in a coiled conformation results in a softening of the multilayers, and consequently in the decrease of the Young modulus. This softening as result of the conformation of the deposited chains is illustrated in the results of Figure 17 where the real, $G'$, and imaginary part, $G''$, of the viscoelasticity shear modulus obtained using a quartz crystal microbalance with dissipation monitoring for (CHI – PAA)$_n$ multilayers assembled under different pH conditions is displayed. The results show that the deposition of layers with the chains in an extended conformation leads to $G' > G''$, with the importance of $G''$ increasing as more coiled chains are deposited [51]. Similar results were reported for the impact of the ionic strength on the mechanical properties of (PDADMAC-PSS)$_n$ multilayers [107].

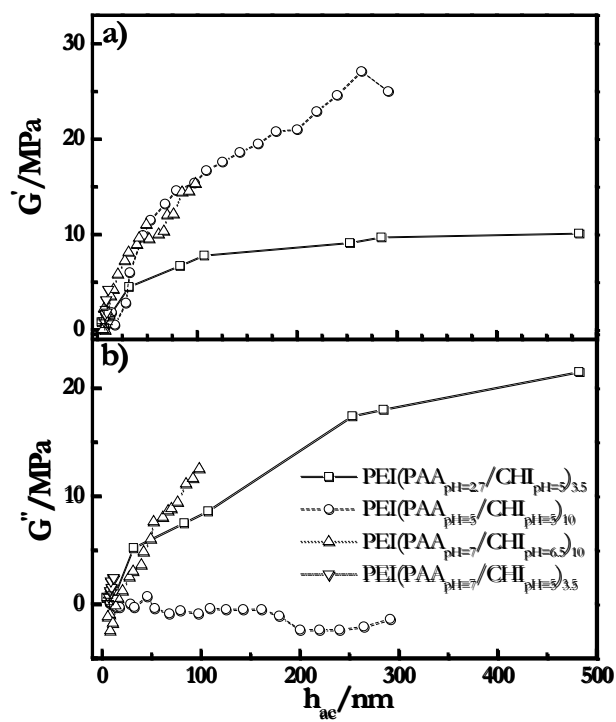

Figure 17. Mechanical properties of the PEI(CHI-PAA)$_n$ multilayer fabricated under different pH conditions as function of the thickness obtained using a dissipative quartz microbalance, $h_{ac}$. (a) $G'$. (b) $G''$. Adapted with permission from Ref. [51]. Copyright (2011) American Chemical Society.

It is worth mentioning that the type of substrates does not alter significantly the mechanical response of polyelectrolyte multilayers [187], with this being generally only affected by the nature of the building blocks and the assembly conditions. This is clear analysing the impact of the increase of the ionic strength used for the assembly of (PDADMAC-PSS)$_n$ multilayers on the mechanical properties of the films. The increase of the ionic strength drives a transition between a purely elastic materials to a fluid-like systems, which is explained in term of the weakening of the ionic pairing as results of a counterion-driven plasticizing. This agrees with the increase of the average water content of the multilayers. Therefore, the modification of any parameter that helps to the increase of the ionic pairing between the layers enhances the rigidity of the materials, whereas any plasticizer effects lead to worsening of this rigidity [51, 107, 187, 337]. This responsiveness can be used for the application of LbL materials in the biomedical field e.g. LbL materials for enhanced cells adhesion [330, 338].

The role of the specific characteristic of the building blocks has been also reported as an aspect allowing the modification of the mechanical properties of the multilayers. Amorim et al. [339] have reported a decrease of the Young's modulus of (PLL-HA)$_n$ multilayers with the increase of molecular weight of the HA used in agreement with the decrease of the hydration of the films. Another possibility for the modification of the mechanical properties of LbL films is the chemical cross-linking of the outermost layers. This enhances the rigidity of the multilayer, and limits the mobility of small molecules through it [104].

### 5.3. Porosity and permeability

Polyelectrolyte multilayers are soft materials, containing a high amount of water. Thus, it seems to be reasonable to analyse two aspects related to the density of the materials: porosity and and permeability. These characteristics play a key role for controlling the exchange of material between the inner region of the films or capsule and the surrounding environment, which present interest in encapsulation process but also for controlling the assembly process and interdiffusion of the polymer chains within the multilayers. It should be expected that rigid multilayers, presenting a high density and as matter of fact a low amount of retained water, results in the formations of non-porous films with a low permeability to molecules along the multi-layered shell.

Several studies have shown that the porosity of LbL films can be tailored at will by controlling carefully the interactions and the choice of the building blocks [121, 332, 335]. This is clear from the different ability of filtration appearing in intrinsically- and extrinsically-compensated films, whereas in the former there is a strong influence on the permeation properties of the capping-layer, the permeability of the latter is governed by the bulk region of the multilayer [340].

Furthermore, the porosity of the obtained materials can be reversibly tuned by changing the ionic strength, pH, temperature, light, ultrasound, magnetic field or mechanical deformation [121]. Antipov et al. [332] showed that the porosity of polyelectrolyte multilayers may be reversibly triggered by changing the pH of the surrounding environment due to the modification of the ionic equilibrium within the multilayer. The selective and tunable permeability of polyelectrolyte multilayers has been recently probed for films of fucoidan and CHI [341]. The chemical nature of the last layer was found to be essential in the control of the swelling degree of the multilayer. Furthermore, the obtained multilayers present size exclusion ability for the penetration of proteins. Thus, whereas lysozyme can permeate and diffuse through the multilayers, bovine serum albumin cannot penetrate into the multilayer. This is indication that specific multilayers may be used for the retention of small proteins and their release upon demand as response to different physico-chemical stimuli.

It is worth mentioning that any factor enabling an enhancement of the film rigidity reduces the permeability of LbL films [121, 342]. This is clear from the studies by Yang et al. [342]. They found that the cross-linking of $(PEI-PAA)_n$ multilayers with glutaraldehyde results in a decrease of the porosity of the films and as matter of fact of their permeability. This agree with the results by Lehaf et al. [331]. They analysed the dependence of the salt diffusion along the multilayer on the mechanical properties of the films and found that the increase of the rigidity was detrimental for ions permeability. Brinke et al. [343, 344] have took advantage of the above characteristic for the fabrication of asymmetric membranes (Chimera membranes), where the deposition of layers with a gradient permeability obtained as results of their assembly under different conditions or even by their different chemical nature, allows the fabrication of materials with enhanced separation properties.

### 5.4. Osmotic response

The discussion of the previous section showed that the modification of the ionic equilibrium during the assembly of polyelectrolyte multilayers impact decisively on the fabrication and properties of the films. This is because the degree of swelling-shrinking of the multilayers is associated with the ionic content of the inner region of the films which affects to the osmotic stress. Therefore, it should be expected that the osmotic response of a preformed LbL material may be associated with any stimuli changing the ionic equilibrium within the film (ionic strength, pH, etc.) [107, 185]. Different studies have pointed out that the modification of the ionic equilibrium in pre-formed multilayers leads to modifications on the films which present sign than that occurring in polyelectrolytes in solution or in polyelectrolyte adsorbing from a solution onto a surface (anti-polyelectrolyte behaviour) [107, 185].

The exposure of (PDADMAC - PSS)$_n$ or (PAH + PSS)$_n$ multilayers to water results in a swelling of the film due to the counterions release which is required to ensure an equality on the chemical potential of the ions inside the multilayer and in the aqueous environment. The opposite phenomena was found when the water is replaced by solution containing salt. This results in the up-taking of counterions for the multilayers which leads to the collapse of the multilayers. It is worth mentioning that this osmotic response, trying to equilibrate the ionic equilibrium, is strongly dependent on the ionic pairing existing within the multilayer and the magnitude of the induced perturbation, presenting in most the cases an almost completely reversibility [107, 185]. Figure 18 shows the dependence of the osmotic response for (PDADMAC-PSS)$_n$ multilayers, obtained as the change on the thickness of the films, after their exposure first to water, and then to a salt solution with the same ionic strength used on the multilayer fabrication. The results evidence that the osmotic shocks is smaller when the modification of the ionic strength is smaller.

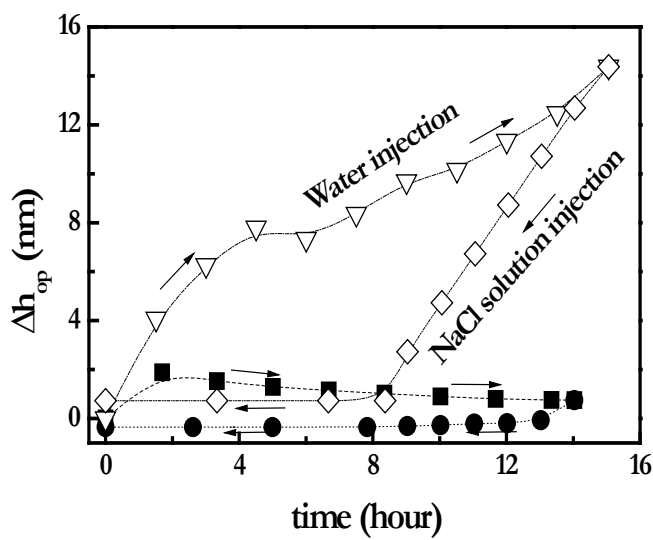

Figure 18. Changes on the thickness measured by ellipsometry as result of a osmotic stresses for (PDADMAC - PSS)$_n$ multilayers built using two different ionic strength. For films built using [NaCl] = 100 mM ■ and ● represent the changes in the thickness when the salt solution is replaced for pure water and the inverse process, respectively and for films built using [NaCl] = 300 mM ∇ and ◊ represent the changes when the salt solution is replaced for pure water and the inverse process, respectively. Adapted from Ref. [107], Copyright (2009), with permission from The Royal Society of Chemistry.

There are some cases in which the introduction of an osmotic stress may result on partial dissolution of the multilayer as was shown by Mjahed et al. [335] for (PLL-HA)$_n$ multilayers. The exposure of such multilayers to ionic strengths above a critical value of 0.3 M results in the formation of holes within the multilayer. The ionic pairing has been stated above as a critical parameter for tuning the swelling degree of polyelectrolytes multilayers, thus it is expected that the swelling can be reduced by a chemical cross-linking between adjacent layers [345]. It is worth noting that the above discussed modifications on the swelling of the multilayers can be a very useful alternative for controlling the release profiles of encapsulated compounds within the multilayer because it affects to the permeability and porosity of the films.

## 6. Applications

The fabrication of both passive and active ("smart materials") nanomaterials by the LbL method and their applications in different areas of the nanotechnology have undergone a spectacular in recent years [30, 31]. The design of such materials take advantage of the possibility to control the film composition along the thickness of the designed supramolecular material, and consequently their properties. Furthermore, the LbL is a simple, fast and cheap approach for the assembly of materials which has fostered the research on its technological applications. However, even though the LbL design of functional materials has been developed as a key enabling technology, the next goal for the LbL technology remains the design and fabrication of 3D ordered nanostructures with real-life applications, which makes it necessary to ensure the scalability of the LbL approach to the industrial level [346]. This is important because the LbL approach provides the bases for controlling the physico-chemical properties of multifunctional materials in such a way which is impossible to attain using other methods such as solution casting, vacuum-assisted filtration or chemical vapor deposition [72]. This section presents a short revision of some of the most promising applications of electrostically assembled LbL materials. It is true that the applications of LbL multilayers based only in polyelectrolytes is rather limited. Therefore, some examples of applications in which at least one of the polyelectrolyte is replaced by other type of building blocks will be discussed.

### 6.1. Layer-by-Layer materials for encapsulation

The fabrication of LbL materials as encapsulation platforms (cargo systems) needs a carefully examination of the specific field of application of the prepared system [41]. The use of the multilayers of LbL deposited onto macroscopic flat templates as cargo systems requires the inclusion of the molecules to be encapsulated either as an independent layer included within the layer stacking or blended with one of the building blocks of the multilayer [50, 51]. This types of material are able to release the encapsulated compound as result of a partial destruction of their structure by a specific external stimuli, e.g. changes in pH, ionic strength, or temperature, photo-dissociation, biodegradation of the material, etc. [347, 348]. Sun et al. [349]used the above discussed approach for the encapsulation and release of different doxorubicin hydrochloride from multilayers of CHI and different polyanions (HA, AG and tannic acid). The degradation of the above multilayers upon their exposure to saline solutions of high pH results in the fast degradation of the multilayers which results in the release of the encapsulated. This can be modify changing the ionic density, molecular structure and functional groups of the polyelectrolytes, with these aspects affecting the morphology, porosity, thickness and physico-chemical properties of the multilayers.

The fabrication of micro- and nano-capsules using the LbL method is a more promising alternative for many technological fields, and in particular for biomedical applications [41, 347]. The research on the applicability of LbL capsules in practical medical situations has received a great interest in recent years [4, 350]. One of the first approaches for the fabrication of capsules by the LbL approach was focused in the assembly of LbL shells onto small drug particles (micro- or nano-meter size drug particles with poor solubility in water). This methodology has been applied for the encapsulation of a broad range materials (anticancer drugs, anticorrosion agents, insoluble dyes, and inorganic salts), enabling a significant enhancement of the dispersion ability of the drug in aqueous medium, e.g. the stability of silver nanoparticles in saline medium was found to be significantly enhanced by the deposition of a LbL shell of CHI and dextran [351]. Furthermore, this type of encapsulation approach ensures a high loading rate of the drugs (70 – 80 wt %). An alternative approach to the above described involves the deposition of multilayers onto sacrificial cores (polymer or ceramic particles) in which the drug may be incorporated [87, 352]. This requires to remove the template once the polyelectrolyte shell is built and before the encapsulation of the molecule. The encapsulation and release of the molecules is attained by controlling the thickness of the multilayer shell and its chemical composition, which can affect to the distribution profile of the encapsulated within the film [353]. This latter approach results in a poorer encapsulation efficiency than the previous one (about 5 – 10 wt %), even though the encapsulation of many chemical using this approach has been possible: ibuprofen, furosemide, nifedipine, naproxen, biotin, vitamin K3, insulin, demathasone, tamoxifen, paclitaxel, and curcumin, DNA fragments, peptides and other therapeutic agents [354, 355]. Selina et al. [356] extended the use of LbL materials on the encapsulation of vaccines with promising results in their preliminary *in vivo* tests using LbL capsules loaded with a plasmid DNA against swine fever. However, the effectiveness of the application of LbL materials as encapsulation platforms, especially for drugs, has associated an additional challenge: the incorporation of specific functionalities favoring molecular recognition processes between the cargo systems and the cells or tissues.

It is worth mentioning that the above discussed methodology for the fabrication of capsules is not limited to the use spherical colloidal particles as templates. Ai et al. [357] fabricated multi-layered polyelectrolyte nanotubes by replication on the inner walls of the pores of alumina templates by alternate deposition of PAH and PSS, followed by the template removal. The obtained nanotubes were found to present a rather thick walls, a high mechanically stability and flexibility. More interesting from the biomedical point of view are the nanotubes of poly(L-arginine) and human serum albumin obtained by Komatsu et al. [358] using nanoporous polycarbonate membranes a template. This type of nanotubes evidenced a high effectiveness in the removal of viral genome from solutions of Hepatitis B virus.

The preparation of LbL capsules requires to consider the role of the interaction between the different templates units in addition to those interactions commonly appearing during the fabrication of LbL materials (assembled block – assembled block, assembled block – solvent and assembled block – template interactions). These template-template interactions may drive the formation of aggregates where the association between the different units can occur through interactions with different strengths, which may induce the fusion or coalescence of soft templates (vesicles/liposomes or emulsion droplets). This may interfere with the assembly process, and reduces the concentration of capsules [42].

The correct protection of the encapsulated compounds requires the deposition of a minimal number layers (commonly around 15 layers) [359]. Additional parameters to consider on the use of LbL materials for encapsulation purpose are the surface charge of the outermost layer, and the degree of swelling and cross-linking of the films, with the latter being correlated to the mechanical properties of the cargo systems. This may modify the release profiles of the encapsulated materials [360]. The control of the above parameters can be performed either during the assembly process or after the multilayer fabrication (post-treatment) by the changing environmental parameters (ionic strength, temperature, pH, solvent quality, etc). This induces a reversible modification on the swelling degree of the capsule which may contribute to the control of the release processes [361]. Antipov et al. [332] showed that the release of encapsulated fluorescein from (PAH - PSS)$_n$ capsules was reduced as the thickness of the multilayer shell increase. This may be rationalized considering the increase of the cross-linking between the layers which leads to a reduction of the permeability of the films. Similar effects were found in relation to the dissolution of calcium oxalate crystals, and the subsequent release of the small inorganic cations formed as result of the decomposition process [362].

The use of LbL materials has solved, at least partially, the difficulties of the control of the release mechanism appearing in most of the traditional systems used for encapsulation. These latter present release profiles mediated by the erosion of the capsule or the free diffusion of the molecules through the shells, which results in most of the case to burst release and poor efficient of the formulations. However, the stimuli responsiveness of LbL materials allow, in many cases, triggering the release using different physical (temperature, light, ultrasounds, magnetic fields, mechanical deformation) or chemical (ionic solvent, pH, solvent quality, electrochemistry) stimuli or even by the living tissues itself, which can be used either for the encapsulation or the controlled release and specific targeting [359, 363, 364]. The type of stimuli used is generally related to the specific nature of the materials, both encapsulated one and those forming the capsule, with the release of the compound being expanded in time-scales in the 1 – 100 hours range. Furthermore, an important advantage of the responsiveness of LbL materials for encapsulation is the possibility to switch on and off the release upon demand [32, 365]. Shen et al. produced (PAH-PSS)$_n$ capsules loaded bovine serum albumin which can be used for loading doxorubicin by the increase of pH of the medium. The release of the drug can be also controlled by the change of the pH due to the modification of the electrostatic interaction between the drug and the capsule [366]. The change of pH was also chosen by De Geest et al. [367] for triggering the release of encapsulated compounds from LbL shells formed by poly(L-arginine) and dextran. They optimize their fabrication process to ensure a release induced by the local pH of the target tissues or cells.

The stimuli responsiveness of LbL multilayers can also be an advantage to facilitate the distribution of the encapsulated compounds towards specific targets. This was used by Podgórna and Szczepanowicz [368] including $Fe_3O_4$ nanoparticles into (PLL-PGA)$_n$ LbL capsules to transport the capsules to an specific target taking advantage of the responsiveness of the particles to magnetic fields. The use of light sensitive materials (mainly to low intensity UV radiation or near infrared radiation) on the fabrication of LbL capsules have also allowed the transport of the nanocapsules to a specific target, which open important possibilities for designing materials with specific targeting abilities [369]. Cheng et al. [370]demonstrated that the application of an electrical field allows triggering the release of gene material (DNA) from its electrostatically assembled multilayers with PEI. The next step for the application of LbL capsules is the fabrication of smart systems enabling the triggering of the release as result of their expose to several stimuli, which provides the bases for an appropriate mimicking of the natural systems [361, 364].

There is a growing interest for the encapsulation of poorly soluble compounds using the liquid environment of oil in water emulsions, where the multi-layered shell provides the protection of the encapsulated compounds, stabilizing the emulsions [359, 371]. This type of encapsulation systems open new perspective for manufacturing nano-containers which can retain the encapsulated compounds during long times, presenting a fast release upon demand [163].

The fabrication of multicapsules or multicompartmental cargo systems (capsosomes) based in the assembly of LbL materials combining multiple subunits such as polyelectrolyte layers, liposomes, and nanoparticles presents also a big interest. The most general approach for preparing this type of multicompartmental systems involves the coating of a colloidal template combining multiple layers of polymers and intact vesicles, which provides the bases for overcoming some of the limitation associated with the application of the individual materials, e.g. the poor mechanical stability of the liposomes. Once the multi-layered structure is formed, the template is removed using the appropriate procedures [98, 99, 101].

## 6.2. Layer by Layer on the fabrication of biocompatible and anti-fouling materials

The versatility of the LbL approach for fabricating materials combining building blocks presenting different physico-chemical properties allows manufacturing materials where the interaction of the materials with cells and tissues is required, including coatings for bone repairing, vascular engineering, tracheal prostheses or dental applications [346, 372-375]. The adhesion, proliferation and differentiation of cells onto polyelectrolyte multilayers is strongly dependent on the film nature, with the multilayer stiffness being critical for their application as substrate for cell culture [376-378]. Furthermore, the use of LbL materials enables the introduction of nutrients, genetic materials or different ligands within the cell culture environment. The use of LbL coatings formed by chitosan and an elastin-like biopolymer onto titanium dental implants was found to be a good alternative for the improvement of the biomineralization and differentiation of osteoblast [379]. LbL coatings formed by 10 bilayers of the PSS-PAH pair onto a polyetheretherketone implant was found to improve the cell adhesion and osseointegration [380]. LbL multilayers can be also useful for improving the cellular proliferation onto cardiovascular implants as was evidenced by Meng et al [381]. They used (CHI – HEP)$_n$ multilayers to coat stainless steel stents, with such coated stent presenting an enhanced affinity for endothelial affinity and thrombus resistance. This facilitates its adhesion to porcine iliac artery endothelial cells and their proliferation onto the coated stent surface (from 20% for the bare metal stent to 60-70% for the coated one).

The antimicrobial properties of implants can be also improved by the use of LbL coatings. The LbL films can face these issue, minimizing the adhesion of the microorganism and killing them by contact or by the release of active compounds close to the implant [382]. Shi et al. [383] showed that the fabrication coatings formed by LbL multi-layered structures of collagen and a cationic antimicrobial peptide on titanium dental implants using allows the reduction of the proliferation of *Staphylococcus aureus* and *Porphyromonas gingivalis*. The deposition of multilayers of lysozyme and collagen onto composite fibers of silk fibroin and nylon 6 was a good alternative to improve the biocompatibility (a significant enhance of the fibroblast proliferation was found in relation to the bare fibers) and mechanical properties of the fiber, reducing the proliferation of *Staphylococcus aureus* and *Escherichia coli* by a 70 and 10%, respectively, in relation to the uncoated fibers [384]. Martins et al. [385] show that multilayers of carrageenan and CHI can be used as scaffolds for tissue engineering, inhibiting the attachment and growth of different bacteria.

The fabrication of antifouling surfaces aimed to the minimization of the adhesion of organic materials is another active field of application of LbL films. It was shown that (PAH - PSS)$_n$ multilayers capped by phosphorylcholine and poly(ethylenoxide) can be used for mimicking the antifouling components of the erythrocyte membranes and limit protein adhesion [386]. Similarly, the combination of LbL polyelectrolyte layers of with a final poly(ethylene-glycol) capping layers was demonstrated to be a good choice for protecting nanolipid carriers loaded with doxorubicing, ensuring their stabilization in the blood stream for long periods of time [387]. Multilayers combining polysaccharides present good properties against the adhesion of serum proteins, and their application onto prosthesis surfaces enhances their thrombi-resistance [374, 388]. This is the result of the high hydration of polysaccharide multilayers. It is worth mentioning that the antifouling properties of LbL materials may be improved by the increase of the density of the films [389].

Etienne et al. [390, 391] show that the growth and proliferation of bacteria onto surface may be prevented using LbL films decorated with defensin and chromofungin. Similarly, Kim et al. [373] pointed out that the proliferation of smooth muscle cell onto stent surfaces may be hindered using multilayers formed by polylysine and hyaluronic acid-grafted-poly(lactic-co-glycolic acid) and loaded with heparin and paclitaxel. The use of multilayers formed by HEP and collagen onto titanium cardiovascular implants was found to inhibit the adhesion and proliferation of platelets, improving the blood compatibility of the metal [392].

### 6.3. Layer-by-Layer materials on the fabrication of membranes

The use of polyelectrolyte multilayers have been frequently for manufacturing membranes for different purposes [31]. LbL films have evidenced a good performance on the selective separation of different pollutant species, e.g. pollutants from water [32, 393-395] or as membranes for pervaporation or ultrafiltration purposes [32, 365]. Furthermore, the good performance of (PAH-PSS)$_n$ multilayers in electroosmotic was evidenced by Qi et al. [396]. They showed that the performance of the membranes depends on both the number of deposited layers and the nature of the outermost layer. Shi et al. [397] showed that polyelectrolyte multilayers of a sulfonated pentablock copolymer and PEI on hydrolyzed polyacrylonitrile fibers may be an very useful tool for dehydrating fuel via pervaporation,

LbL films are a promising tool for the fabrication of membraned for pressure-driven desalination, mainly membranes for reverse and forward osmosis [398]. Zhang et al. [399] showed that porous membranes coated by a multilayer of polyvinylamine and polyvinylsulfate presented a good performance in both reverse and forward osmosis, with its rejection of $MgCl_2$ and $MgSO_4$ being almost independent of concentration of the feed solution or the operating pressure. However, the permeation of other salts, such as NaCl and $Na_2SO_4$, was found to be strongly dependent on the operational parameters. However, even though the above membranes present under specific conditions good salt rejection, the water flux is low. An important drawback to the use of LbL films in desalination process is that they are easily destroyed under severe conditions, including high ionic strengths or chlorine treatments. This makes it necessary to improve their stability by cross-linking processes [400, 401]. Qiu et al. [400] showed that the use of membranes formed by PAH and PSS cross-linked with glutaraldehyde leads to a significant enhancement of the $MgCl_2$ in relation to their non-cross-linked counterpart. However, the water permeability of the rejection layers was significantly reduced.

Qi et al. [396] showed that the dependence of the properties of the capping layer can influence decisively on the application of (PAH-PSS)$_n$ in the fabrication of forward osmosis membranes, allowing a controlled modification of the hydraulic permeability and the solute permeability. This was explained in terms of a complex interplay of interactions between the electrolytic solutions and multilayer. Furthermore, the efficiency of the membrane was found to be easily controllable by tuning the number of deposited layers or the chemical nature of the capping layer.

Lee et al. [402]took advantage on the LbL approach for fabricating (PAH-PAA)$_n$ multilayers loaded with methylene blue and heparin onto electrospun polyacrylonitrile fibers. The obtained membranes showed good antifouling and anticoagulation properties, i.e. good blood compatibility. This makes them a good alternative for their use as biomedical membranes for hemodialysis.

Additional applications of the membranes obtained using the LbL method are the fabrication of chiral membranes for separation of optical active compounds [372], membranes for selective ion separation [376] or membranes for the optimization of the proton transport in fuel cells [377].

**6.4. Layer-by-Layer materials for the fabrication of self-healing and anti-corrosion coatings**

The LbL method has been developed as a powerful tool on the fabrication of self-healing materials because it allows the introduction of nano-reservoirs or nano-reactors within the multilayer structure which may provide a self-healing character to the manufactured materials [403-405]. The activation of the healing characteristics of nano-reservoirs is generally triggered by external stimuli with physical, chemical or mechanical origin. Shchukin et al. [406] showed that polyelectrolyte multilayers with self-healing properties may protect to aluminium surfaces against corrosion.

Moreover, polyelectrolyte multilayers can present intrinsic self-healing properties, which may be related to the dynamics of the polyelectrolytes chains within the multilayer, i.e. the mobility of the chains in the hydrated environment of the multilayer can help on the healing of the films after a surface damage [404]. The self-healing character of (PEI-PAA)$_n$ multilayers triggered by the water penetration in the internal regions of the films was observed by Wang et al. [407], with the water penetration favoring the interdiffusion of the polyelectrolytes from the internal layers to the outermost region of the multilayers. However, the self-healing of polyelectrolyte multilayers depend on their structure and thickness, with the existence a strong cross-linking in the structure reducing the self-healing properties of the films as result of a hindered interdiffusion of the chains. This allows rationalizing the limited self-healing properties of (PDADMAC - PSS)$_n$ multilayers in term of the strong interactions between the polyelectrolyte chains once the multilayers is formed [408], which limits the self-healing on polyelectrolyte multilayers to those involving weak polyelectrolytes [404]. Recently, Yuan et al. [409] prepared LbL multilayers using a peptide modified carboxymethyl chitosan and dopamine modified oxidized alginate (OALG-D). These multilayers evidenced a significant cell adhesion for human dermal fibroblast, holding high self-healing and radical scavenging abilities. These characteristics make the multilayers a promising alternative for regenerative medicine applications.

The LbL multilayers have been also applied on the fabrication of anticorrosion protective coatings. Farhat and Schlenoff [410] showed that the fabrication of a (PDADMAC-PSS)$_n$ multilayer with ten bilayers onto stainless steel surfaces allows a significant reduction of the metal corrosion, with the anticorrosion properties being improved for intrinsically compensated multilayers where the permeability is significantly reduced due to the strong ionic pairing between polyelectrolytes in adjacent layers. The fabrication of multilayers containing hydrophobic polyelectrolytes - (poly(N-octadecyl-2-ethynyl-pyridinum bromide)-poly(ether-ether-ketone sulfonate))$_n$ or (poly(vinyl-pyridine)-PSS)$_n$- are a good alternative for anticorrosion protection [410].

Udoh et al. [411] tested, very recently, the ability of multilayers formed by different combinations of polyelectrolytes loaded with mesoporous silica particles containing benzotriazole as self-healing protective anticorrosion coatings for aluminum alloys, and found that multilayers formed by two weak polyelectrolytes release the healing compounds faster than multilayers combining weak and strong polyelectrolytes, with the later allowing for a more controlled and prolonged release. Furthermore, it was found that the anticorrosion properties was strongly dependent on the nature of the multilayer and the number of polyelectrolyte layers.

## 6.5. Layer-by-Layer materials in the fabrication of superhydrophobic/superhydrophilic coating

The fabrication of superhydrophobic and superhydrophilic surfaces by the LbL approach requires the combination of polyelectrolytes and nanoparticles to control the roughness of the films, and minimize the contact angle hysteresis [236, 412]. (PAH-PAA)$_n$ multilayers with embedded silica nanoparticles and capped with a fluorinated copolymer allow fabricating superhydrophobic coatings with a contact angle around of 175º [413]. The inclusion of ZrO$_2$ particles on (PAH-PAA)$_n$ multilayers allows fabricating surfaces with a good water repellency. This is strongly correlated to the number of deposited (at least 20 bilayers are needed for an optimal repellency) and the nature of the last layer (PAH-capped multilayers evidenced better repellency) [414]. Han et al. [415] showed that multilayers the wettability properties of the multilayers combining PAH and silica nanoparticles can be reversibly modified from superhydrophobic (contact angle>165º) to superhydrophilic (contact angle<10º) by exposure to UV-ozone plasma.

Huang et al. [416] shows that textured (PAH-PAA)$_n$ multilayers allows the manufacturing of omniphobic and slippery (slip angle for water and oils around 3), with the type of texture being tuned by change of pH. (PAH-PAA)$_n$ multilayers were also used by Guo et al. [417] to create self-cleaning surfaced as were shown by. They coated electrospun polyacrylonitrile fibers with this type of multilayers, and found that the coated fibers present a good ability to separate oil-water mixtures and emulsions with high flux and oil recovery efficiency under intermittent pressure. This is due to their superamphiphobic properties. Thus the obtained structures present both air superhidrophilicity and a complete oil repellency when they are wetted by water i.e. underwater superoleophobicity. Furthermore, the coated fibers have excellent self-cleaning properties.

It is worth mentioning that the above discussion on the possible applications of LbL materials in different technological fields is not intended to be exhaustive and only selected cases have been chosen. The list of possible applications of LbL materials also include energy storage devices, chemical and biological sensors or high-strength composite films, to cite some examples [4, 6, 72, 418]. However, the detailed discussion of the multiple potential applications of LbL materials may result in several reviews, and remains far from the scope of this review.

## 7.   Concluding remarks

This review has examined some of the most fundamental bases of the Layer-by-Layer method for the fabrication of multi-layered systems, paying special attention to the analysis of the building process, properties and potential applications of systems involving polyelectrolytes, the so-called polyelectrolyte multilayer. This is a rapidly evolving field which difficulties the task of presenting a comprehensive description of these systems, hence this work tries to provide a general perspective of the current knowledge on this broad field.

The simplicity, flexibility and versatility of the LbL method have resulted in a strong development, which leads to a continuous appearance of new concepts, procedures and applications enabling the fabrication of functional nanomaterials with a broad range of properties and structures. This has required a careful examination of the complex physico-chemical bases underlying the assembly process. The formation of polyelectrolyte multilayers is possible through a complex interplay of different contributions -electrostatic interactions (enthalpic contribution) *vs.* entropic contributions- that determines the ionic pairing, the growth and structure of the obtained multilayers, and their physico-chemical properties. It is clear that only from the understanding of the interactions involved in the assembly process is possible to understand the particular characteristics of this type of systems, which help on the control of the distribution of the polyelectrolyte during the fabrication of the multilayer and consequently of the multilayer structure (stratification of the layers). The control of such aspect opens many routes to the application of polyelectrolyte multilayers in different industrial and technological fields. It is true that the use of the LbL approach on the fabrication of functional materials may be a promising starting point for the nano-architectonics, i.e. the combination of the nanotechnological concepts with other scientific fields (organic chemistry, supramolecular chemistry, and biotechnology) for the fabrication of materials. However, the extensive research efforts made in the understanding of the physico-chemical bases of the building of polyelectrolyte multilayers cannot hide the lack of knowledge remaining in some particular aspects, especially related to the adsorption dynamics, which is an important drawback for the development of real applications of these systems.

## Acknowledgements

This work was funded by MINECO under grant CTQ2016-78895-R and by Banco Santander-Universidad Complutense grant PR87/19-22513. We are grateful to C.A.I. Espectroscopia from the UCM for the use of their facilities.